\documentclass{IEEEoj}

\usepackage[T1]{fontenc}
\usepackage{color,array,amsthm}
\usepackage{graphicx}
\usepackage{cite}
\usepackage{amsmath,amssymb,amsfonts}
\usepackage{float}
\usepackage{CJK}
\usepackage{textcomp}
\usepackage{xcolor}
\usepackage{colortbl}
\usepackage{url}
\usepackage{multirow}
\usepackage{graphicx}
\usepackage{hyperref}
\usepackage{subfigure}
\usepackage{wrapfig}
\usepackage{latexsym}
\usepackage{algorithm}
\usepackage{algpseudocode}

\def\BibTeX{{\rm B\kern-.05em{\sc i\kern-.025em b}\kern-.08em
    T\kern-.1667em\lower.7ex\hbox{E}\kern-.125emX}}
\AtBeginDocument{\definecolor{ojcolor}{cmyk}{0.93,0.59,0.15,0.02}}
\def\OJlogo{\vspace{-14pt}\includegraphics[height=28pt]{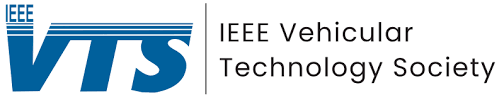}}

\begin{document}
\receiveddate{XX November,2023}
\reviseddate{XX Month, 2023}
\accepteddate{XX Month, 2023}
\publisheddate{XX Month, 2023}
\currentdate{2 November,2022}
\doiinfo{OJVT.2023.1234567}

\title{Free-Space Optical (FSO) Satellite Networks Performance Analysis: Transmission Power, Latency, and Outage Probability} 

\author{JINTAO LIANG (Member, IEEE)\textsuperscript{1}}
\author{AIZAZ U. CHAUDHRY (Senior Member, IEEE)\textsuperscript{1}}
\author{EYLEM ERDOGAN (Senior Member, IEEE)\textsuperscript{2}}
\author{HALIM YANIKOMEROGLU (Fellow, IEEE)\textsuperscript{1}} 
\author{GUNES KARABULUT KURT (Senior Member, IEEE)\textsuperscript{3}} 
\author{PENG HU (Senior Member, IEEE)\textsuperscript{4}}
\author{KHALED AHMED\textsuperscript{5}}
\author{STEPHANE MARTEL\textsuperscript{5}}
\affil{Department of Systems and Computer Engineering, Carleton University, Ottawa, ON K1S 5B6, Canada}
\affil{Department of Electrical and Electronics Engineering, Istanbul Medeniyet University, Istanbul, 34700, Turkey}
\affil{Poly-Grames Research Center, Department of Electrical Engineering, Polytechnique Montreal, Montreal, QC H3T 1J4, Canada}
\affil{National Research Council of Canada (NRC), Ottawa, ON K1A 0R6, Canada}
\affil{Technology Strategy of Satellite Systems, MDA, Sainte-Anne-de-Bellevue QC H9X 3Y5, Canada}
\authornote{This work was supported in part by the High Throughput and Secure Networks (HTSN) Challenge Program of the National Research Council of Canada (NRC), and in part by MDA.}
\markboth{Preparation of Papers for IEEE OPEN JOURNALS}{Author \textit{et al.}}

\begin{abstract}
In free-space optical satellite networks (FSOSNs), satellites can have different laser inter-satellite link (LISL) ranges for connectivity. As the LISL range increases, the number of satellites from among all the satellites in the constellation that will be needed on the shortest path between a source and a destination ground station decrease, and thereby the number of the LISLs on the shortest path decreases. Greater LISL ranges can reduce network latency of the path but can also result in an increase in transmission power for satellites on the path. Consequently, this tradeoff between satellite transmission power and network latency should be investigated, and in this work we examine it in FSOSNs drawing on the Starlink Phase 1 Version 3 (i.e., the latest version of Starlink’s Phase 1) and Kuiper Shell 2 (i.e., Kuiper’s biggest shell) constellations for different LISL ranges and different inter-continental connections. We use appropriate system models for calculating the average satellite transmission power (i.e., the average of the transmission power of all satellites on the shortest path) and network latency (i.e., the end-to-end latency of the shortest path). The results show that the mean network latency (i.e., the mean of network latency over all time slots) decreases and mean average satellite transmission power (i.e., the mean of average satellite transmission power over all time slots) increases with an increase in LISL range. For the Toronto--Sydney inter-continental connection in an FSOSN with Starlink's Phase 1 Version 3 constellation, when the LISL range is approximately 2,900 km, the mean network latency and mean average satellite transmission power intersect are approximately 135 ms and 380 mW, respectively. For an FSOSN with the Kuiper Shell 2 constellation in this inter-continental connection, this LISL range is around 3,800 km, and the two parameters are approximately 120 ms and 700 mW, respectively. For the Toronto--Istanbul and Toronto--London inter-continental connections, the LISL ranges at the intersection are different and vary from 2,600 km to 3,400 km. Furthermore, we analyze outage probability performance of optical uplink/downlink due to atmosphere attenuation and turbulence. \\
\end{abstract}



\maketitle

\section{Introduction}
Free-space optical wireless communications are being developed rapidly and have a promising future. In particular, free-space optical satellite networks (FSOSNs) \cite{b1} will be realized through laser inter-satellite links (LISLs) \cite{b2}, also known as optical inter-satellite links, in upcoming low Earth orbit (LEO) satellite constellations, such as SpaceX’s Starlink \cite{b3},\cite{b4}, Amazon’s Kuiper \cite{b5}, and Telesat’s Lightspeed \cite{b6}. These new FSOSNs will have various advantages compared to satellite networks using radio frequency based inter-satellite links (ISLs), including larger link bandwidth, license-free spectrum, higher link data rate, and lower terminal power required to establish ISLs. On the other hand, FSOSNs have some downsides, such as attenuation of the optical signal during transmission through Earth’s atmosphere \cite{b7} and a limitation on the amount of energy their satellite solar panels can obtain \cite{b8}. As the mission lifetime of a satellite is mainly determined by its battery life, maximizing a satellite's battery life is critical for minimizing the replacement cost of the satellites. Since battery life is determined by the number of charging and recharging cycles, if the satellite consumes less power, the battery recharge time will decrease, and its lifetime can increase. The analysis of the power consumption by the communication system of the satellite with respect to the total power budget of the satellite is out of the scope of this paper. However, satellite transmission power can be an important factor that can affect battery life and, by extension, the lifetime of the satellite \cite{b9}. Therefore, in this work, we focus on the analysis of satellite transmission power for optical links.

For LISLs in FSOSNs, the optical beam from a transmitting satellite can be received without attenuation. For uplink (i.e., the link between a source ground station and an ingress satellite in the FSOSN) and downlink (i.e., the link between an egress satellite in the FSOSN and a destination ground station), we consider optical links instead of radio frequency links, since optical links provide larger link bandwidth, higher link data rate, better security, smaller aperture size, and lower terminal power consumption compared to radio frequency links \cite{b1},\cite{b10}. For optical uplink and downlink, the optical beams must propagate through Earth’s atmosphere to reach the receiver, and multiple attenuation affect the beams and thereby attenuate the transmitted optical signal. Thus, satellite transmission power for optical uplink/downlink must consider the losses caused by the atmosphere, such as those due to Mie scattering and geometrical scattering \cite{b11}. In addition, it's essential to consider atmospheric turbulence, as it has the potential to induce signal fading and consequently lead to a deterioration in the quality of the communication, including outage probability.

Latency is an important parameter to gauge the performance of data communications over satellite networks in various applications, such as the high-frequency trading of stocks between financial stock exchanges around the world \cite{b12}, information exchanges for military purposes \cite{b13}, etc. In such communications networks, there are four components of network latency, namely propagation delay, transmission delay, queuing delay, and processing delay. Propagation delay only depends on the propagation distance between the source and destination over the satellite network, while the rest are affected by the link data rate, the switching and routing speed, and the traffic status \cite{b14}. The processing delay, transmission delay, and queuing delay constitute the node delay. 

In FSOSNs, satellites have an LISL range for connectivity, which is the range within which a satellite can successfully establish an LISL with another satellite. For longer LISL ranges, optical links between satellites on the shortest path over an FSOSN have longer propagation distances, and higher satellite transmission power is required to maintain reliable links between satellites. On the other hand, fewer satellites are needed to establish the shortest path for an inter-continental connection between two ground stations over an FSOSN, resulting in less node delay and thereby less network latency of the shortest path. In this way, different LISL ranges lead to a tradeoff between satellite transmission power and network latency, which is the focus of this paper. 

To study this tradeoff, we employ the Starlink Phase 1 Version 3 \cite{b4} and Kuiper Shell 2 \cite{b5} constellations, and we consider three different inter-continental connections, specifically Toronto to Sydney, Toronto to Istanbul, and Toronto to London, and different LISL ranges for satellites. The results show that as the LISL ranges increases, the mean of network latency (i.e., the mean of network latency of the shortest paths at all time slots for an inter-continental connection) decreases. At the same time, the mean of average satellite transmission power (i.e., the mean of average satellite transmission power of the shortest paths at all time slots for an inter-continental connection) increases, where the average satellite transmission power is the mean of the transmission power of all satellites on a shortest path at a time slot. 

In addition to the observation through figures, the results further show that for the Toronto--Sydney inter-continental connection over the Starlink Phase 1 Version 3 constellation, the mean of network latency and mean average satellite transmission power intersect when the LISL range is approximately 2,900 km and are approximately 135 ms and 380 mW, respectively. For the Kuiper Shell 2 constellation in this inter-continental connection, this LISL range is around 3,800 km, and these two parameters are approximately 120 ms and 700 mW, respectively. For the Toronto--Istanbul inter-continental connection, the two parameters intersect when LISL ranges are about 2,600 km for Starlink Phase 1 Version 3 constellation and 2,900 km for Kuiper Shell 2 constellation. For the Toronto to London inter-continental connection, these parameters intersect when LISL ranges are 3,400 km and 3,000 km for the Starlink Phase 1 Version 3 and Kuiper Shell 2, respectively. Based on the results, we find that for LISL and laser uplink/downlink, the network latency for the shortest path for an inter-continental connection between two ground stations and the average satellite transmission power for the shortest path have an inverse relationship. For different LISL ranges for the two FSOSNs over different inter-continental connections, we find that there is no clear relationship between the LISL range when network latency and satellite transmission power reach balance and the length of the inter-continental connection within the same FSOSN. Furthermore, we find that the performance of the outage probability is inevitably influenced by cloud conditions.

The rest of the paper is organized as follows. Related work and motivation are discussed in Section II. Section III presents the system model, including models for laser link transmission power, models for turbulence induced fading channel, and models for link and node latency. Section IV introduces the two constellations and provides methodology to calculate LISL and ground station ranges, orbital period, outage probability, network latency, and average satellite transmission power. Section V discusses the results for the tradeoff between average satellite transmission power and network latency, the effect of atmosphere turbulence on outage probability, and mentions some insights based on these results. Conclusions and directions for future work are provided in Section VI. 

\section{Related Work and Motivation}
Several studies have discussed the relationship between satellite power consumption and battery life and have analyzed latency for satellite networks, such as \cite{b8}, \cite{b9}, \cite{b15,b16,b17,b18,b19,b20,b211,b221,b21,b22} and the references therein.

In \cite{b8}, the authors discussed satellite battery lifetime and indicated that typical satellite missions in a LEO system are projected to last for 3 to 5 years, which can be roughly regarded as a range of 16,500 to 27,500 partial cycles. In order to maintain such a large number of cycles, battery lifetime and approaches to reduce the satellite power consumption should be considered.

In \cite{b9}, the authors conducted a study on power management in Low Earth Orbit (LEO) satellite constellations, emphasizing the critical role of the electrical power system in ensuring the Quality of Service (QoS) for LEO satellite communications. They developed a model for satellite battery characteristics based on various communication traffic scenarios and proposed several approaches to extend the battery lifetime.

In \cite{b15}, the authors highlighted the constraints of limited satellite power and stressed the need for resource allocation mechanisms to take satellite power consumption during transmission into account, particularly for downlink communication. Due to the increasing importance of on-board transmission power optimization, they introduced an innovative carrier and power allocation method for satellite systems to reduce satellite power consumption.

In \cite{b16}, the authors introduced a communication methodology designed to efficiently manage the communication resources of LEO satellites in order to extend the battery lifetime. They proposed a system design aimed at  extending the battery lifetime on board the satellites. This system can enable the deployment of large-scale LEO satellite constellations while satisfying the communication requirements of current satellite communications. 

In \cite{b17}, the authors focused on the power consumption required for satellite terminal communications as well as the battery Depth of Discharge (DOD) during the communication period. They also discussed power control systems, and the results of their study showed that with the reduction of the transmission power, less DOD is required per cycle and therefore longer battery life can be achieved and, by extension, longer satellite lifetimes.

In \cite{b18}, the authors simulated Starlink Phase 1 Version 1 constellation and established their study based on the fundamental characteristics of the Starlink deployment. They conducted simulations of routing designs within this satellite network. They analyzed the constellation parameters and offered several preliminary insights into the dynamics of the constellation's evolving topology. They also discussed the possible routing strategies and the resultant end-to-end latency properties of this constellation.

In \cite{b19}, the authors framed the problem of inter-satellite topology design for
large LEO constellation, including Starlink and Kuiper. They proposed a new approach for designing satellite networks using regular repetitive patterns and investigated the topology and impact of ISL range and setup time of satellite network based on this approach. Furthermore, they evaluated and compared the network performance of the two satellite constellations. 

In \cite{b20}, the authors studied the link reliability of the software defined satellite network (SDSN). They designed a failure probability model for ISLs that considered queuing and processing delays and the corresponding SDSN reliability analysis was added. The authors simulated the satellite network with certain constellation and latency parameters, including queuing delay and processing delay, and analyzed the network performance.

In \cite{b211}, the author investigated the latency performance of large LEO constellations. The author studied the use of ground-based relays for the LEO constellations to provide low-latency wide area networking. For upcoming mega-constellations like Starlink, the author mentioned two inter-continental connections and the results indicated that implementing ISLs could reduce latency and supplement capacity in heavy traffic in network.

In \cite{b221}, the authors investigated the time delay and outage performance for inter-satellite transmissions, satellite-terrestrial transmissions, terrestrial relay assisted transmissions, and satellite relay assisted transmissions for LEO mage-constellations. They proposed analytical models to investigate the end-to-end time delay and outage performance while considering different transmission scenarios. They presented numerical results to validate the proposed analytical models and compare the performances in terms of time delay and outage probability. They further concluded that the size of coverage area of LEO satellites exhibited a weak negative influence on time delay.

In \cite{b21}, the authors focused on the reliability performance of downlink transmission in LEO satellite communication. They derived a novel expression for the outage probability of LEO satellite transmission. They simulated the LEO satellite communication scenarios with severe Doppler effects and analyzed the performance of the system based on orthogonal time frequency space (OTFS) scheme and traditional orthogonal frequency division multiplexing (OFDM) scheme.

The tradeoff between average satellite transmission power and network latency in FSOSNs arises from different LISL ranges. Figs. 1 and 2 show different shortest paths for different LISL ranges for the same inter-continental connection from Toronto to Istanbul. In these figures, uplinks and downlinks are highlighted in green, while LISLs are shown in yellow. In Fig. 1, satellites have a longer LISL range and there are only three satellites on the shortest path. In this way, each LISL is longer, and each satellite needs higher transmission power to an LISL. However, only three node delays are added to the network latency in addition to propagation delays of the four links. In Fig. 2, satellites have a shorter LISL range, and so there are more satellites on the shortest path, which leads to higher network latency but shorter LISLs and therefore lower satellite transmission power. 

In an earlier work in \cite{b7}, we studied system models for transmission power and link margin for the LISL and laser uplink/downlink to examine the impact of link distance and link margin on LISL transmission power, and the impact of slant distance, elevation angle, and link margin on laser uplink/downlink transmission power. In another earlier work in \cite{b22}, we proposed a binary integer linear program formulation of the total network latency over multiple LISL ranges and multi-pairs of inter-continental connection and simulated the Starlink Phase 1 Version 3 constellation. Furthermore, we considered several satellite transmission power constraints to investigate the tradeoff. In this work, we use these models from \cite{b7} and \cite{b22} in calculating the average satellite transmission power for satellites on the shortest path between a source and destination ground station over an FSOSN for studying the tradeoff between average satellite transmission power and network latency in FSOSNs. Investigations on satellite transmission power and latency have been conducted separately in the literature, but a tradeoff between satellite transmission power and network latency and analysis of the outage probability has never been investigated before. We consider both transmission delay and node delay while analyzing the tradeoff, and we obtain the results based on simulation of the on-going Starlink Phase 1 Version 3 and Kuiper Shell 2 constellation. To the best of our knowledge, this is the first study, which focuses on the tradeoff between average satellite transmission power and network latency for FSOSNs with the Starlink Phase 1 Version 3 and Kuiper Shell 2 constellations, and investigate the outage probability due to atmosphere turbulence. This is the first work of its kind, and no similar previous works exist.

\begin{figure*}[htbp]
\centerline{\includegraphics[width=16cm, height=4.5cm]{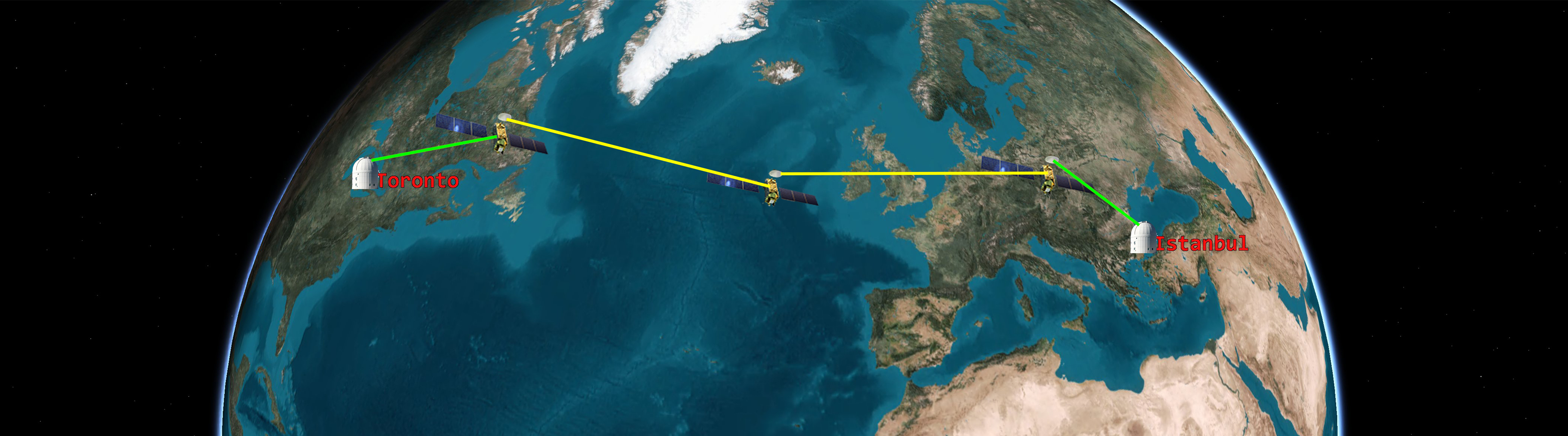}}
\caption{An illustration of a shortest path for a longer LISL range for an inter-continental connection from Toronto to Istanbul.}
\label{fig}
\end{figure*}

\begin{figure*}[htbp]
\centerline{\includegraphics[width=16cm, height=4.5cm]{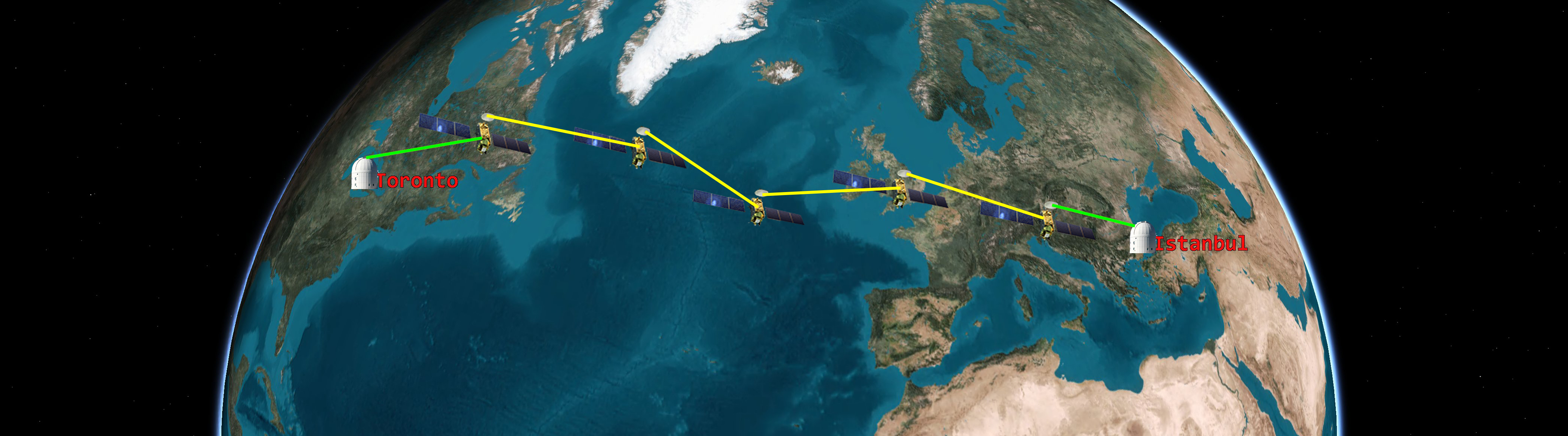}}
\caption{An illustration of a shortest path for a shorter LISL range for an inter-continental connection from Toronto to Istanbul.}
\label{fig}
\end{figure*}

\section{System Model}
Here, we introduce the system model for FSOSNs, which includes models for LISL transmission power, laser uplink/downlink transmission power, link margin, turbulence induced fading channel, link latency, and node latency.

\subsection{Transmission Power in FSOSNs}
In the following, we introduce the laser link transmission power model and the link margin model. 

\subsubsection{Laser inter-satellite link transmission power model} 
For the inter-satellite link, the transmission power $\textit{P}_T$ is given by \cite{b23}
\begin{equation}
\label{eq_1}
\textit{P}_T = \textit{P}_R/(\textit{G}_T\textit{G}_R\textit{L}_T\textit{L}_R\textit{L}_{PS}\textit{$\eta$}_T\textit{$\eta$}_R) \text{,}
\end{equation}
where $\textit{P}_T$ is the transmitted power for LISL in Watts, $\textit{P}_R$ is the received power for the link in Watts, $\textit{G}_T$ is the transmitter gain, $\textit{G}_R$ is the receiver gain, $\textit{L}_T$ is the transmitter pointing loss, $\textit{L}_R$ is the receiver pointing loss, $\textit{L}_{PS}$ is the free-space path loss for LISL, $\textit{$\eta$}_T$ is the transmitter optics efficiency, and $\textit{$\eta$}_R$ is the receiver optics efficiency. 
The transmitter gain $\textit{G}_T$ in (1) is given by \cite{b24}
\begin{equation}
\label{eq_2}
\textit{G}_T = 16/({\textit{$\Theta$}_T})^{2}\text{,}
\end{equation}
where $\textit{$\Theta$}_T$ is the full transmitting divergence angle in radians. The receiver gain $\textit{G}_R$ in (1) is given by \cite{b23}
\begin{equation}
\label{eq_3}
\textit{G}_R = (\textit{$\pi$}\textit{D}_R/\textit{$\lambda$})^2 \text{,}
\end{equation}
where $\textit{D}_R$ is the receiver’s telescope diameter in mm, and $\textit{$\lambda$}$ is the operating wavelength in nm. The transmitter pointing loss $\textit{L}_T$ in (1) is given by \cite{b23} 
\begin{equation}
\label{eq_4}
\textit{L}_T = \exp{(-\textit{G}_T(\textit{$\theta$}_T)^{2})}\text{,}
\end{equation}
where $\textit{$\theta$}_T$ is the transmitter pointing error in radians. The receiver pointing loss $\textit{L}_R$ in (1) is given by \cite{b23} 
\begin{equation}
\label{eq_5}
\textit{L}_R = \exp{(-\textit{G}_R(\textit{$\theta$}_R)^{2})}\text{,}
\end{equation}
where $\textit{$\theta$}_R$ is the receiver pointing error in radians. The free-space path loss for LISL $\textit{L}_{PS}$ is given by \cite{b23}
\begin{equation}
\label{eq_6}
\textit{L}_{PS} = (\textit{$\lambda$}/4\textit{$\pi$}\textit{d}_{SS})^2\text{,} 
\end{equation} 
where $\textit{$\lambda$}$ is the operating wavelength in nm, and $\textit{d}_{SS}$ is the propagation distance between satellites in km.
\vspace{1ex}

\subsubsection{Laser uplink/downlink transmission power model}
Uplink and downlink are the links between a ground station and satellite. In this work, we consider Mie scattering and geometrical scattering in terms of atmospheric attenuation for optical uplink/downlink. Mie scattering can be caused by the particles in the medium and it can redirect the signal from its intended direction, while geometrical scattering causes reflection, refraction, and scattering of the optical signal.

Atmospheric attenuation must be considered to model laser uplink and downlink, and the transmission power for laser uplink/downlink $\textit{P}_T$ is given by \cite{b25}
\begin{equation}
\label{eq_7}
\textit{P}_T = \textit{P}_R/(\textit{G}_T\textit{G}_R\textit{L}_T\textit{L}_R\textit{L}_A\textit{L}_{PG}\textit{$\eta$}_T\textit{$\eta$}_R) \text{,}
\end{equation}
where $\textit{L}_A$ is the atmospheric attenuation loss, $\textit{L}_{PG}$ is the free-space path loss for laser links between ground stations and satellites, and the rest of the parameters are same as the LISL transmission power model in (1). The slant distance (i.e., the distance between a ground station and a satellite) for laser uplink/downlink $\textit{d}_{GS}$ is given by \cite{b26}
\vspace{0.17cm} 
\begin{equation}
\label{eq_8}
\textit{d}_{GS} = \textit{R}(\sqrt{((\textit{R}+ \textit{H})/\textit{R})^2-(\cos({\textit{$\theta$}_E}))^{2}}-\sin({\textit{$\theta$}_E}))\text{,}
\end{equation} 
where $\textit{R}$ $=$ $\textit{r}$ $+$ $\textit{h}_E$, $\textit{H}$ $=$ $\textit{h}_S$ $-$ $\textit{h}_E$, $\textit{r}$ is the radius of the Earth, $\textit{h}_E$ is the height of the ground station, $\textit{h}_S$ is the altitude of the satellite, and $\textit{$\theta$}_E$ is the elevation angle of the ground station in degrees. The free-space path loss $\textit{L}_{PG}$ can be expressed on the basis of slant distance accordingly:
\begin{equation}
\label{eq_9}
\textit{L}_{PG} = (\textit{$\lambda$}/4\textit{$\pi$}\textit{d}_{GS})^2\text{.} 
\end{equation}

\textit{a) Atmospheric attenuation due to Mie scattering}\\
\hspace*{1em}
When the diameter of atmosphere particles equals or exceeds the wavelength of the optical beam, it results in Mie scattering. This occurs because the phase of the optical beam is non-uniform across the particle and leads to the deviation of the transmitted signal from its intended direction. It is mainly caused by microscopic particles of water, and it happens in the lower part of the atmosphere where these particles are more abundant. Mie scattering can be modeled by the following expression \cite{b27}:
\begin{equation}
\label{eq_10}
\textit{$\rho$} = \textit{a}(\textit{h}_E)^{3}+\textit{b}(\textit{h}_E)^{2}+\textit{c}\textit{h}_E+\textit{d}\text{,}
\end{equation} 
where $\textit{$\rho$}$ is the extinction ratio, $\textit{h}_E$ is the height of the ground station in km, and $\textit{a}$, $\textit{b}$, $\textit{c}$ and $\textit{d}$ are empirical coefficients dependent on wavelength, which can be expressed as follows \cite{b27}:
\begin{equation}
\label{eq_11}
\textit{a} = -0.000545\textit{$\lambda$}^2+0.002\textit{$\lambda$}-0.0038\text{,}
\end{equation}
\begin{equation}
\label{eq_12}
\textit{b} = 0.00628\textit{$\lambda$}^2-0.0232\textit{$\lambda$}+0.00439\text{,}
\end{equation}
\begin{equation}
\label{eq_13}
\textit{c} = -0.028\textit{$\lambda$}^2+0.101\textit{$\lambda$}-0.18\text{,}
\end{equation}
\begin{equation}
\label{eq_14}
\textit{d} = -0.228\textit{$\lambda$}^3+0.922\textit{$\lambda$}^2-1.26\textit{$\lambda$}+0.719\text{.}
\end{equation}
\hspace*{1em}The atmospheric attenuation due to Mie scattering can be calculated as
\begin{equation}
\label{eq_15}
\textit{I}_m = \exp{(-\textit{$\rho$}/\sin({\textit{$\theta$}_E}))}\text{,} 
\end{equation} 
where $\textit{$\theta$}_E$ is the elevation angle of the ground station in degrees.
\vspace{1ex}

\textit{b) Atmospheric attenuation due to geometrical scattering}\\
\hspace*{1em}
Geometrical scattering is used to model the attenuation that takes place near the surface of the Earth and is produced by fog, dense clouds, water molecules, rain or other atmospheric phenomena. Geometrical scattering can be modeled by the following expression \cite{b11}:
\begin{equation}
\label{eq_16}
\textit{V} = 1.002/(\textit{N}\textit{L}_W)^{0.6473} \text{,} 
\end{equation} 
where $\textit{V}$ is the visibility in km, $\textit{N}$ is the cloud number concentration in cm\textsuperscript{-3}, and $\textit{L}_W$ is the liquid water content in g/m\textsuperscript{-3}. The attenuation coefficient $\textit{$\theta$}_A$ can be expressed as \cite{b11}
\begin{equation}
\label{eq_17}
\textit{$\theta$}_A = (3.91/\textit{V}) (\textit{$\lambda$}/550)^{-\textit{$\varphi$}}\text{,} 
\end{equation}
where $\textit{$\varphi$}$ is the particle size related coefficient based on Kim’s model. The Beer-Lambert law is given by \cite{b28}
\begin{equation}
\label{eq_18}
\textit{I}(\textit{z})= \exp{(-\textit{$\mu$}\textit{z})}\text{,} 
\end{equation}
where $\textit{$\mu$}$ is the wavelength dependent attenuation coefficient, and $\textit{z}$ is the distance of the transmission path. The atmospheric attenuation caused by geometrical scattering can be expressed using the Beer-Lambert law as
\begin{equation}
\label{eq_19}
\textit{I}_g= \exp{(-\textit{$\theta$}_A\textit{d}_A)}\text{,} 
\end{equation} 
where $\textit{d}_A$ is the distance of the laser beam through the troposphere and can be calculated as \cite{b7}
\begin{equation}
\label{eq_20}
\textit{d}_A= (\textit{h}_A-\textit{h}_E)\csc({\textit{$\theta$}_E})\text{,} 
\end{equation}
where $\textit{h}_E$ is the height of the ground station in km, $\textit{h}_A$ is the height of the troposphere in km, and $\textit{$\theta$}_E$ is the elevation angle of the ground station in degrees.\\
\hspace*{1em} In this work, we consider geometrical scattering for uplink, and Mie scattering and geometrical scattering for downlink. The atmospheric attenuation loss for uplink considering only  geometrical scattering can be expressed as 
\begin{equation}
\label{eq_21}
\textit{L}_{Aup} = \textit{I}_g = \exp{(-\textit{$\theta$}_A\textit{d}_A)}\text{,} 
\end{equation}
and the atmospheric attenuation loss for downlink considering both Mie scattering and geometrical scattering can then be expressed as 
\begin{equation}
\label{eq_211}
\textit{L}_{Adown} = \textit{I}_m\textit{I}_g = \exp{(-\textit{$\rho$}/\sin({\textit{$\theta$}_E}))}\exp{(-\textit{$\theta$}_A\textit{d}_A)}\text{.} 
\end{equation}

\subsubsection{Link Margin Model}
The link margin is defined as the ratio of the received signal power $\textit{P}_R$ to the required signal power $\textit{P}_{req}$ to achieve a specific bit error rate (BER) at a given data rate, and it can be represented as \cite{b25}
\begin{equation}
\label{eq_22}
\textit{LM} = \textit{P}_R/\textit{P}_{req}\text{.} 
\end{equation}
\hspace*{1em} Since we are interested in computing the link transmission power in this work, we need to know the received power based on link margin. The required signal power is generally referred to as receiver sensitivity, and the received power can be calculated as
\begin{equation}
\label{eq_23}
\textit{P}_R = \textit{LM}×\textit{P}_{req}\text{.} 
\end{equation} 

\subsection{Atmosphere turbulence in FSOSNs}
In the following, we introduce the an atmosphere channel model which is under the influence of
atmosphere turbulence resulting in intensity fluctuations. 

In this work, we consider that the turbulence induced channel experiences an exponentiated Weibull fading channel.The probability density function (PDF) is given by \cite{b255}
\begin{multline}
\label{eq_233}
\textit{f}(\textit{I}) = 
(\textit{$\alpha$}\textit{$\beta$}/\textit{$\eta$})(\textit{I}/\textit{$\eta$})^{\textit{$\beta$}-1}\exp{(-(\textit{I}/\textit{$\eta$})^{\textit{$\beta$}})}\\
(1-\exp{(-(\textit{I}/\textit{$\eta$})^{\textit{$\beta$}})})^{\textit{$\alpha$}-1} \text{,} 
\end{multline}
and the cumulative distribution function (CDF) is given by \cite{b255}
\begin{equation}
\label{eq_234}
\textit{F}(\textit{I}) = (1-\exp{-(\textit{I}/\textit{$\eta$})^{\textit{$\beta$}}})^{\textit{$\alpha$}} \text{,} 
\end{equation}
where $\textit{$\alpha$}$, $\textit{$\beta$}$ are the shape parameters and $\textit{$\eta$}$ is the scale parameter of a ground station. The three parameters can be expressed as \cite{b256}
\begin{equation}
\label{eq_235}
\textit{$\alpha$} = 7.220\textit{$\alpha$}^{2/3}/\Gamma{(2.487\textit{$\sigma$}_I^{1/3}-0.104)} \text{,} 
\end{equation}
\begin{equation}
\label{eq_236}
\textit{$\beta$} = 1.012(\textit{$\alpha$}\textit{$\sigma$}_I^2)^{-0.52}+0.142 \text{,} 
\end{equation}
\begin{equation}
\label{eq_237}
\textit{$\eta$} = 1/(\textit{$\alpha$}\Gamma{(1+1/\textit{$\beta$})}\textit{g}(\textit{$\alpha$},\textit{$\beta$})) \text{,} 
\end{equation}
where $\textit{g}(\textit{$\alpha$},\textit{$\beta$})$ is the $\textit{$\alpha$}$ and $\textit{$\beta$}$ dependent constant variable, and $\textit{$\sigma$}_I$ is the scintillation index of the ground station, which is expressed as \cite{b257}
\begin{multline*}
\label{eq_4}
\sigma_I = \exp\left( 0.49 \sigma_R^2 / (1 + 1.11 \sigma_R^{2.4})^{7/6} \right. \\
+ \left. 0.51 \sigma_R / (1 + 0.69 \sigma_R^{2.4})^{5/6} \right) - 1,
\end{multline*}
where the Rytov variance $\textit{$\sigma$}_R$ is given by \cite{b257}
\begin{equation}
\label{eq_239}
\textit{$\sigma$}_R = 2.25\textit{k}^{7/6}\sec^{11/6}{(\textit{$\zeta$})}\int_{\textit{h}_E}^{\textit{h}_S} \textit{C}_n^2(\textit{h})(\textit{h}-\textit{h}_E)^{5/6} \,dh \text{,}
\end{equation}
where \textit{k} is the optical wave number, $\textit{$\zeta$}$ is the zenith angle of the ground station, $\textit{h}_S$ is the altitude of the satellite, and $\textit{C}_n^2$ is the altitude dependent refractive index constant, which is expressed as \cite{b258}
\begin{multline}
\label{eq_240}
\textit{C}_n^2(\textit{h}) = 8.148\times10^{-56}\textit{v}_r^2\textit{h}^{10}\textit{e}^{-\textit{h}/1000}\\
+2.7\times10^{-16}\textit{e}^{-\textit{h}/1500}+\textit{C}_0\textit{e}^{-\textit{h}/100}\textit{m}^{-2/3} \text{,}
\end{multline}
where $\textit{v}_r$ is the r.m.s. ground wind speed, and $\textit{C}_0$ is the nominal value of the refractive index constant at ground.

\subsection{Latency in FSOSNs}
In the following, we introduce the modeling of latency in FSOSNs. In optical communication networks, four types of delay can be considered as part of the end-to-end latency. These delays consist of propagation delay, transmission delay, queueing delay, and processing delay. Propagation delay of an optical link depends on its propagation distance, which we also refer to as link latency. Node delay, also referred to as node latency in this work, comprises transmission, queueing, and processing delays at a satellite.

\subsubsection{Link Latency Model}
The propagation delay of a link (or link latency) is the delay the transmitted signal takes to be received by the receiver through the channel. It depends on the propagation medium and distance. In FSOSNs, the propagation delay of a laser link $\textit{L}_{p}$ or link latency $\textit{T}_{link}$ can be calculated as 
\begin{equation}
\label{eq_24}
\textit{T}_{link} = \textit{L}_{p} = \textit{d}_{link}/\textit{c}_s\text{,} 
\end{equation}
where $\textit{d}_{link}$ is the length of the laser link (or propagation distance between transmitter and receiver satellites), and $\textit{c}_s$ is the speed of light, which is 299,792,458 m/s \cite{b29}.

\subsubsection{Node Latency Model}
According to \cite{b1}, satellite required a precise acquisition, tracking, and pointing (ATP) system for effective and correct connection with other satellites due to the narrow beam divergence of the laser beam and the different satellite motion velocities. Based on the parameters of the using optical terminals and routers, current setup delays for satellites to complete ATP process to establish temporary LISLs for nodes switching are usually as huge as a few seconds and prohibitive for LISLs. Therefore, in this work we consider the node delay without setup delay.\\
\hspace*{1em} Transmission delay refers to the delay it takes to transmit a packet on the channel on the transmitter side. It depends on the size of the data packet and the data rate of the link. In FSOSNs, transmission delay $\textit{L}_{t}$ can be calculated as
\begin{equation}
\label{eq_25}
\textit{L}_{t} = \textit{l}/\textit{R}_{data}\text{,} 
\end{equation}
where $\textit{l}$ is the length of a packet, and $\textit{R}_{data}$ is the data rate of the laser link. Transmission delay can be minimal if the link data rate is very high. For example, if the link data rate is 10 Gbps and the packet size is 1,500 bytes, $\textit{L}_{t}$ is 1.2 $\mu$s, which is negligible.\\
\hspace*{1em} Queuing delay $\textit{L}_{q}$ refers to the delay that a packet incurs while waiting in the buffer of a router. If a new packet arrives while an old one is still waiting or being transmitted, it will stay in the buffer and wait. For an $\textit{M}$/$\textit{M}$/1 queuing system \cite{b30}, where the first $\textit{M}$ denotes the packet arrival rate, the second $\textit{M}$ denotes the packet service rate, and 1 denotes the number of servers in the queuing system, the queuing delay $\textit{L}_{q}$ is discussed in \cite{b31}. The authors assumed a service rate of 740 kpps (where kpps stands for kilo packets per second), an arrival rate of 726 kpps, a link data rate of 9 Gbps, and a packet size of 1,500 bytes with an extra 25 bytes of header and footer and calculated $\textit{L}_{q}$ to be 4 ms. Based on this calculation, we assume the queuing delay as 4 ms for 10 Gbps data rate communication.\\
\hspace*{1em}The processing delay $\textit{L}_{proc}$ is the time taken by a router to process a packet to make an appropriate routing and switching decision before sending the packet to the next appropriate hop. According to \cite{b20}, when a router has enough processing capability, it can be assumed that the router has approximately equal queuing and processing delays for simplification, and the authors assumed queuing delays and processing of 4 ms and 6 ms, respectively, in their simulations. For LISLs, it's reasonable to assume that the processing as 6 ms under the assumptions of a congestion-free FSOSN having LISLs with very high data rates as 10 Gbps.\\
\hspace*{1em} In this work, we consider the node delay or node latency $\textit{T}_{node}$ as the sum of queuing and processing delays, since the transmission delay is negligible for a 10 Gbps link data rate, and we assume a node delay of 10 ms for each satellite in the FSOSN, i.e., 
\begin{equation}
\label{eq_26}
\textit{T}_{node} = \textit{L}_{q} + \textit{L}_{proc} = 10 \text{ ms}\text{.} 
\end{equation}
\hspace*{1em} The LEO satellites in upcoming satellite constellations, such as Starlink, Kuiper, and Telesat, will have the packet switching and routing capability \cite{b3,b4,b5,b6}, and the queuing and processing delays will indeed be investigated in the node latency model.

\section{Methodology}
In this section, we describe the Starlink Phase 1 Version 3 and Kuiper Shell 2 constellations and present the methodology for the calculation of the LISL range, ground station range, shortest path for an inter-continental connection over the FSOSN, network latency, and average satellite transmission power.

\subsection{Overview of Starlink and Kuiper Constellations}
We introduce the two satellite constellations, i.e., Starlink Phase 1 Version 3 \cite{b4} and Kuiper Shell 2 \cite{b5} constellations, as described in their FCC filings as of 2020 and 2019, respectively. For a satellite constellation, its constellation orbit parameters including number of orbital planes, number of satellites on each plane, and inclination angle can also affect network latency. Greater number of orbital planes and satellites on each plane lead to denser distribution of satellites within the constellation, thus more satellites can be connected for the same LISL range, thereby more optimal links established to achieve the shortest path over an inter-continental connection and results less propagation delay added to the network latency. Similarly, less inclination angle leads to denser satellite distribution in the space above the areas away from the pole, and less propagation delay for inter-continental connections across the low latitude areas.

\subsubsection{Starlink Phase 1 Version 3 Constellation}
The latest version of the Starlink Phase 1 constellation contains 1,584 satellites and has 22 orbital planes with 72 satellites in each plane. Its altitude is 550 km, and its inclination is 53°. We assume the phasing parameter as 17 to prevent intra-constellation satellite collisions as per \cite{b32}. The Walker notation for this constellation is 53°:1,584/22/17, and Fig. 3 shows the constellation pattern. Table 1 provides parameters for different versions of the Starlink Phase 1 constellation since its original proposal in 2016.

\begin{figure}[htbp]
\centerline{\includegraphics[width=8cm, height=6cm]{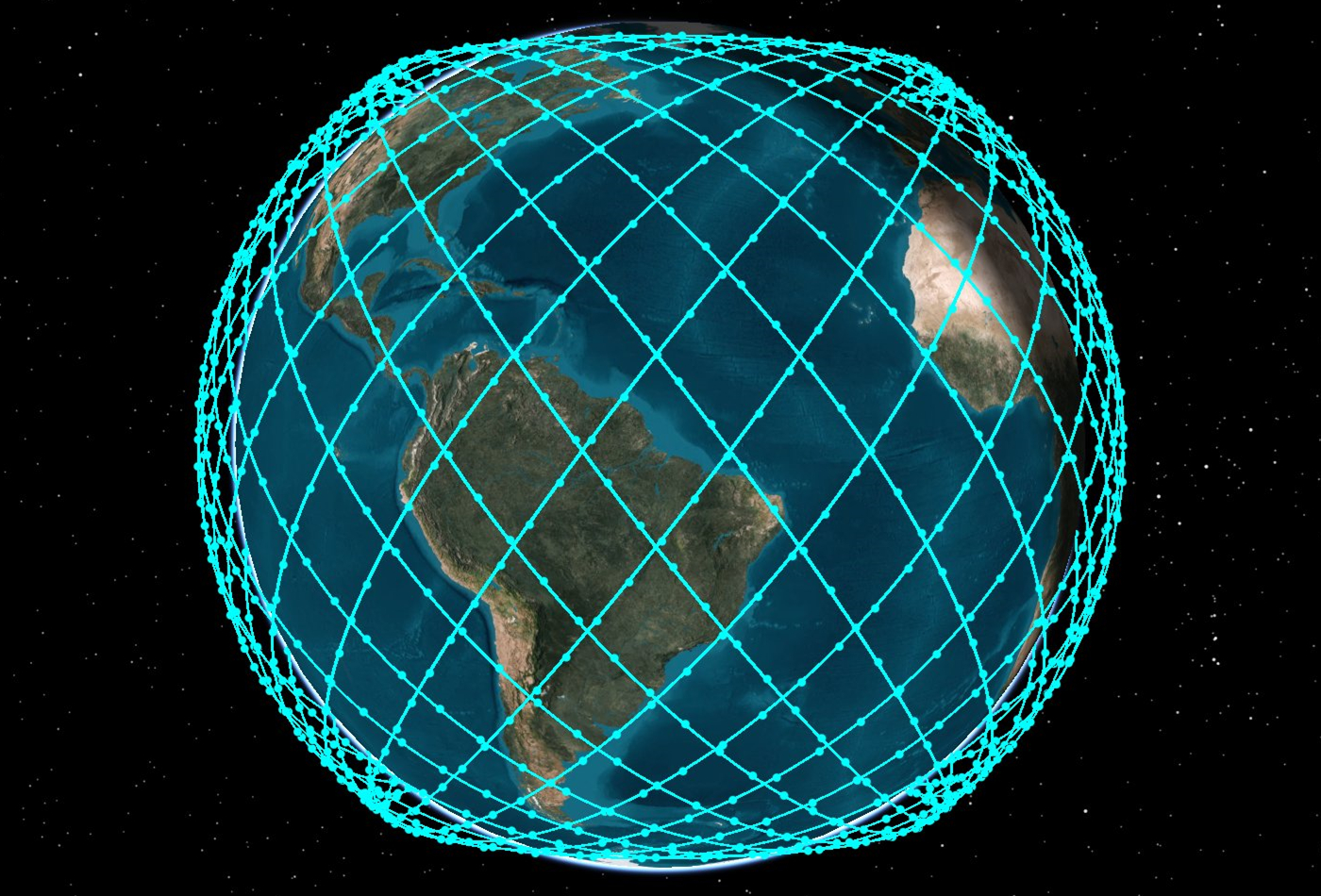}}
\caption{Starlink phase 1 version 3 constellation.}
\label{fig}
\end{figure}

\begin{table}
\centering
\renewcommand\thetable{1}
\caption{Summary of Parameters for Different Versions of the Starlink Phase 1 Constellation.}
\arrayrulecolor{black}
\begin{tabular}{!{\color{black}\vrule}l!{\color{black}\vrule}l!{\color{black}\vrule}l!{\color{black}\vrule}l!{\color{black}\vrule}} 
\hline
\textbf{Parameter} & \begin{tabular}[c]{@{}l@{}}\textbf{Version 1} \\ \textbf{\cite{b31}}\end{tabular} & \begin{tabular}[c]{@{}l@{}}\textbf{Version 2} \\ \textbf{\cite{b3}} \end{tabular} & \begin{tabular}[c]{@{}l@{}}\textbf{Version 3}\\ \textbf{\cite{b4}} \end{tabular} \\ 
\hline
Number of Orbital Planes & 32 & 24 & 22 \\ 
\hline
\begin{tabular}[c]{@{}l@{}}Number of Satellites \\ per Plane\end{tabular} & 50 & 66 & 72 \\ 
\hline
Total Satellites & 1,600 & 1,584 & 1,584 \\ 
\hline
Altitude (km) & 1,150 & 550 & 550 \\ 
\hline
Inclination (°) & 53 & 53 & 53 \\
\hline
\end{tabular}
\arrayrulecolor{black}
\end{table}

\subsubsection{Kuiper Shell 2 Constellation}
The second shell of the Kuiper constellation—Kuiper Shell 2—is the biggest of its three shells. This constellation contains 1,296 satellites in total with 36 satellites distributed uniformly on each of the 36 orbital planes. The altitude of this constellation is 610 km, and the inclination is 42°. Based on the analysis in \cite{b32}, the most suitable value of the phasing parameter for intra-constellation satellite collision avoidance is 11 for Kuiper Shell 2 constellation. Its Walker constellation is 42°:1,296/36/11, and Fig. 4 shows the constellation pattern. Table 2 presents the main parameters of the different shells of the Kuiper constellation \cite{b5}.

\begin{figure}[htbp]
\centerline{\includegraphics[width=8cm, height=6cm]{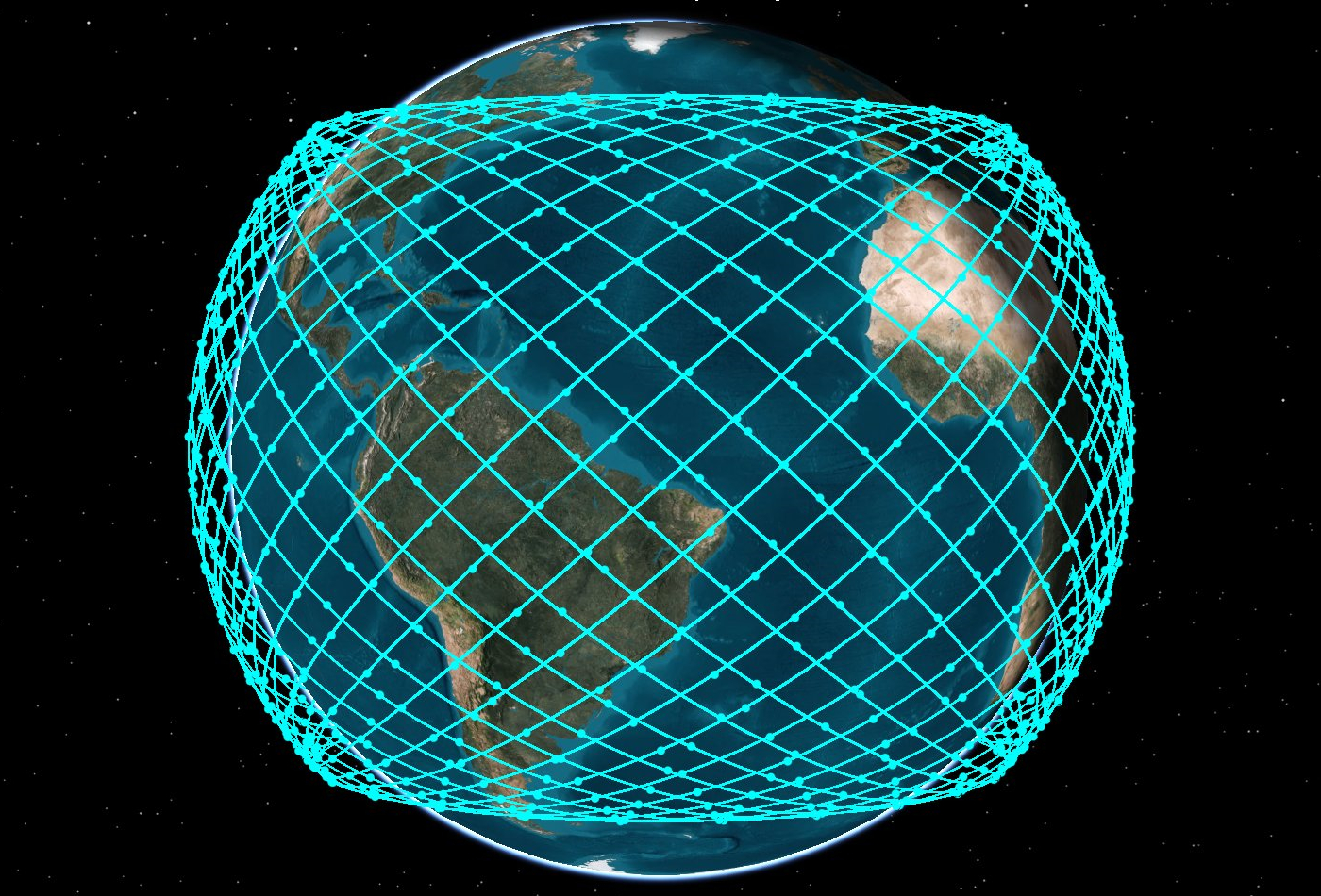}}
\caption{Kuiper shell 2 constellation.}
\label{fig}
\end{figure}

\begin{table}
\centering
\renewcommand\thetable{2}
\caption{Summary of Parameters for Different Shells of Kuiper Constellation.}
\arrayrulecolor{black}
\begin{tabular}{!{\color{black}\vrule}l!{\color{black}\vrule}l!{\color{black}\vrule}l!{\color{black}\vrule}l!{\color{black}\vrule}} 
\hline
\textbf{Parameter} & \textbf{Shell 1} & \textbf{Shell 2} & \textbf{Shell 3} \\ 
\hline
Number of Orbital Planes & 28 & 36 & 34 \\ 
\hline
Number of Satellites per Plane & 28 & 36 & 34 \\ 
\hline
Total Satellites per Shell & 784 & 1,296 & 1,156 \\ 
\hline
Altitude (km) & 590 & 610 & 630 \\ 
\hline
Inclination (°) & 33 & 42 & 51.9 \\
\hline
\end{tabular}
\arrayrulecolor{black}
\end{table}

\subsection{LISL Range Calculation}
To study the tradeoff between the average satellite transmission power and network latency in FSOSNs, the LISL range is a critical factor to consider. The LISL range affects the connectivity of satellites, since a satellite can only establish LISLs with other satellites that are within this range. The maximum LISL range for satellites in a constellation can be regarded as the range which is constrained only by visibility, and it can be calculated as \cite{b2}
\begin{equation}
\label{eq_27}
\textit{D}_{S} = 2{\sqrt{(\textit{r}+ \textit{h}_S)^2 - (\textit{r}+ \textit{h}_a)^2}}\text{,}
\end{equation} 
where $\textit{r}$ is the radius of the Earth, which is generally assumed to be 6,378 km, $\textit{h}_S$ is the altitude of the satellite, $\textit{h}_{a}$ is the height of the Earth's atmosphere, which is generally considered to be 80 km, and $\textit{D}_{S}$ is the maximum LISL range. For the Starlink Phase 1 Version 3 constellation, $\textit{h}_S$ is 550 km, and the maximum LISL range can be calculated as 5,016 km. Similarly, for the Kuiper Shell 2 constellation with an altitude of 610 km, the maximum LISL range is 5,339 km. \\
\hspace*{1em} To study the tradeoff between network latency and average satellite transmission power, we consider several LISL ranges for satellites in FSOSNs for the two constellations. For the Starlink Phase 1 Version 3 constellation, we assume different LISL range values, specifically 1,575 km, 1,731 km, 2,000 km, 3,000 km, 4,000 km, and 5,016 km. The 1,575 km LISL range is a reasonable minimum feasible LISL range for a satellite in this constellation, as a satellite can establish six permanent LISLs with other satellites at this range, including four with neighbors in the same orbital plane and two with its nearest left and right neighbors in adjacent orbital planes. Shorter LISL ranges are not sufficient for a satellite to establish the two permanent LISLs with neighbors in adjacent orbital planes. At the 1,731 km LISL range, a satellite can establish ten permanent LISLs with other satellites, including four with neighbors in the same orbital plane and six with its nearest neighbors in adjacent orbital planes. \\
\hspace*{1em} For the Kuiper Shell 2 constellation, we also consider different LISL ranges, in this case 1,515 km, 2,000 km, 3,000 km, 4,000 km, and 5,339 km. Due to its different design, the Kuiper Shell 2 constellation has a different orbital plane distribution and a different satellite distribution per orbital plane than the Starlink Phase 1 Version 3 constellation; a satellite in the former constellation needs a longer LISL range to establish permanent LISLs with neighbors in same orbital plane as compared to the LISLs with nearest neighbors in the adjacent orbital planes. At the 1,515 km LISL range, a satellite in the Kuiper Shell 2 constellation can set up eight permanent LISLs, including two with satellites in the same orbital plane and six with satellites in adjacent orbital planes. For the Kuiper Shell 2 constellation, 1,515 km is a reasonable minimum LISL range for satellites to establish a reasonable minimum connectivity, since shorter LISL ranges are not long enough for a satellite to establish permanent LISLs with neighbors in same orbital plane.

\subsection{Ground Station Range Calculation}
The connectivity for the uplink and downlink is affected by the range of the ground station, as the greater the ground station range, the longer the propagation distance for uplink and downlink. The distance of the link between a ground station and satellite can be calculated using (8). The uplink/downlink is longest when the elevation angle between the ground station and an ingress/egress satellite is at its minimum, and the maximum range of the ground station can be calculated for the minimum elevation angle for each constellation. For the Starlink Phase 1 Version 3 constellation, the minimum elevation angle is 25°, and the corresponding maximum range of a ground station is 1,123 km. For the Kuiper Shell 2 constellation, the minimum elevation angle is 35°, and the maximum ground station range is 1,412 km.

\subsection{Satellite Orbital Period Calculation}
Network latency and average satellite transmission power should be studied based on the entire orbital period of satellites in an FSOSN, which is the time a satellite takes to complete one full orbit around the Earth. The orbital period of a satellite in an FSOSN can be calculated using \cite{b34} 
\begin{equation}
\label{eq_28}
\textit{T} = 2\textit{$\pi$}{\sqrt{(\textit{r}+ \textit{h}_S)^3 / (\textit{G}\textit{M}_{E})}}\text{,}
\end{equation} 
where $\textit{r}$ is the radius of the Earth, which is 6,378 km, $\textit{h}_S$ is the altitude of the satellite in the FSOSN, $\textit{G}$ is the gravitational constant, which is $6.673\times10^{-11}$ Nm\textsuperscript{2}/kg\textsuperscript{2}, and $\textit{M}_{E}$ is the mass of the Earth, which is $5.98\times10^{24}$ kg. Based on the altitude of the Starlink Phase 1 Version 3 and Kuiper Shell 2 constellations, which are 550 km and 610 km, respectively, the orbital period for a satellite in these constellations is 5,736 seconds and 5,810 seconds, respectively.

\subsection{Outage Probability Calculation}
The outage probability (OP) is defined as the probability that the signal-to-noise (SNR) $\textit{$\gamma$}_{th}$ can fall below a certain threshold with acceptable communication quality. 
The OP in this case can be defined as 
\begin{equation}
\label{eq_281}
\textit{P}_{out} = Pr\{\textit{$\gamma$}\leq\textit{$\gamma$}_{th}\} \text{,}
\end{equation}
Substituting (26) into (38), the OP can be expressed as \cite{b43}
\begin{equation}
\label{eq_282}
\textit{P}_{out} = (1-\exp{(-(\textit{$\gamma$}_{th}/(\textit{$\eta$}\textit{L}_a)^2)\bar{\textit{$\gamma$}})^{\textit{$\beta$}/2}})^\textit{$\alpha$} \text{,}
\end{equation}
where $\bar{\textit{$\gamma$}}$ is the average SNR, and $\textit{L}_a$ is the atmosphere attenuation loss for uplink/downlink.
In this work, we investigate the impact of cloud condition on the atmosphere attenuation and turbulence, and therefore the outage probability of optical uplink/downlink communication. We assume three cloud condition scenarios, the first one is clear weather condition with a thin cirrus cloud concentration of 0.5 cm\textsuperscript{-3} and liquid water content as $3.128\times10^{-4}$ g/m\textsuperscript{-3}, while the second one is clear weather condition with a cirrus cloud concentration of 0.025 cm\textsuperscript{-3} and liquid water content as 0.06405 g/m\textsuperscript{-3}, and the third one is cloudy weather condition with a Cumulus cloud concentration of 250 cm\textsuperscript{-3} and liquid water content as 1.0 g/m\textsuperscript{-3}, according to \cite{b10}. Based on these three different weather conditions, we calculated the OP using practical parameters and we hope these deployment scenarios can be used to provide practical insights about the system architecture, so that the reader can get a quick idea about the overall system performance.

\subsection{Shortest Path Calculation}
To study the tradeoff between the average satellite transmission power and network latency in an FSOSN at different LISL ranges, we need to calculate the shortest path between the source ground station and the destination ground station over the FSOSN at an LISL range at each time slot to calculate the network latency and average satellite transmission power for that shortest path. The shortest path can be determined based on adjacency matrix containing propagation distances of laser links between satellites and between satellites and ground stations using Dijkstra’s algorithm \cite{b35}. This algorithm uses weights of the links to determine the shortest path, and these weights are the propagation distances of the optical links in our case. 
The following steps are used in Dijkstra’s algorithm to find the shortest path for an inter-continental connection from a source ground station to a destination ground station over an FSOSN. \\
\indent 1) Create a shortest path set $\textit{S}$ that keeps track of satellites that are on the shortest path, another set $\textit{X}$ for satellites that are not on the shortest path, and a vector $\textit{D}$ to track the shortest path distance for each satellite to the source ground station $\textit{g}_{s}$. \\
\indent 2) Assign an initial weight to all the links as their propagation distance, add $\textit{g}_{s}$ to $\textit{S}$, and update the distance for $\textit{g}_{s}$ in $\textit{D}$ as 0, while the distances for other satellites to $\textit{g}_{s}$ are infinite in $\textit{D}$. \\
\indent 3) While $\textit{S}$ does not include the destination ground station $\textit{g}_{d}$, go through the following loop. \\ 
\indent a) Pick the satellite $\textit{q}$ with the minimum distance to $\textit{g}_{s}$, which is the closest satellite to $\textit{g}_{s}$ based on its distance to $\textit{g}_{s}$. \\
\indent b) Mark $\textit{q}$ as visited and include $\textit{q}$ to $\textit{S}$. Update the distance for $\textit{q}$ in $\textit{D}$. \\
\indent c) For every adjacent satellite $\textit{p}_{i}$ of $\textit{q}$, check its distance to $\textit{q}$. Since Dijkstra’s algorithm keeps tracking the currently known shortest path for each satellite from the source ground station and updates the distance if it finds a shorter one, check the new path to $\textit{g}_{s}$ through $\textit{q}$ from $\textit{p}_{i}$. For each $\textit{p}_{i}$, if the currently known distance of $\textit{p}_{i}$ to $\textit{g}_{s}$ is greater than the sum of the distance of $\textit{q}$ to $\textit{g}_{s}$ and the distance between $\textit{q}$ and $\textit{p}_{i}$, then update the distance for $\textit{p}_{i}$ to $\textit{g}_{s}$. After iterating through all the adjacent satellites $\textit{p}_{i}$ of $\textit{q}$, go back to step $\textit{a}$ to find the next satellite to add to $\textit{S}$.

\subsection{Network Latency Calculation}
We calculate the network latency $\textit{T}_{net}$ of a shortest path at a time slot as the end-to-end latency of the shortest path between the source ground station and the destination ground station by adding the link latency (or propagation delay) of each laser link, including the uplink, downlink, and ISLs, and the node delay (or node latency) for each satellite on the shortest path, which can be expressed as
\begin{equation}
\label{eq_29}
\textit{T}_{net} = (\textit{T}_{up} + \sum_{k=1}^{n-1} \textit{T}_{k} + \textit{T}_{down}) + \textit{n}×\textit{T}_{node}\text{,}
\end{equation}
where $\textit{T}_{up}$ and $\textit{T}_{down}$ are the link latencies of the laser uplink and downlink, respectively, and $\textit{T}_{k}$ is the link latency (or propagation delay) for \textit{k}th ISL. Since there are \textit{n} satellites on the shortest path for the inter-continental connection, $\textit{n}-1$ ISLs exist in total on the shortest path. As mentioned in Section III, $\textit{T}_{node}$ is the node delay for each satellite and is assumed to be 10 ms. 

The propagation delay for each link on the shortest paths for first five time slots for the FSOSN with the Starlink Phase 1 Version 3 constellation at the 3,000 km LISL range for the Toronto--Sydney inter-continental connection is shown in Table 3. $\textit{T}_{up}$ and $\textit{T}_{down}$ represent the propagation delay for the uplink and downlink, and $\textit{T}_{k}$ is the propagation delay for the \textit{k}th ISL. For example, the network latency of the shortest path at the first time slot is equal to the sum of $\textit{T}_{up}$, $\textit{T}_{down}$, the propagation delays of all seven LISLs, and $\textit{T}_{node}$ for each satellite on the shortest path (i.e., 20 ms × 8), and is equal to 137.22 ms.
\begin{table*}
\centering
\renewcommand\thetable{3}
\caption{Propagation Delay for Each Link on the Shortest Paths at the First Five Time Slots for the FSOSN with the Starlink Phase 1 Version 3 Constellation at 3,000 km LISL Range for the Toronto--Sydney Inter-Continental Connection.}
\arrayrulecolor{black}
\begin{tabular}{!{\color{black}\vrule}l!{\color{black}\vrule}l!{\color{black}\vrule}l!{\color{black}\vrule}l!{\color{black}\vrule}l!{\color{black}\vrule}l!{\color{black}\vrule}l!{\color{black}\vrule}l!{\color{black}\vrule}l!{\color{black}\vrule}l!{\color{black}\vrule}l!{\color{black}\vrule}} 
\hline
\begin{tabular}[c]{@{}l@{}}\textbf{Time} \\ \textbf{Slot} \end{tabular} & \begin{tabular}[c]{@{}l@{}}\textbf{$\textit{T}_{up}$} \\ \textbf{(ms)} \end{tabular}& \begin{tabular}[c]{@{}l@{}}\textbf{$\textit{T}_{1}$} \\ \textbf{(ms)} \end{tabular}& \begin{tabular}[c]{@{}l@{}}\textbf{$\textit{T}_{2}$} \\ \textbf{(ms)} \end{tabular} & \begin{tabular}[c]{@{}l@{}}\textbf{$\textit{T}_{3}$} \\ \textbf{(ms)} \end{tabular} & \begin{tabular}[c]{@{}l@{}}\textbf{$\textit{T}_{4}$} \\ \textbf{(ms)} \end{tabular} & \begin{tabular}[c]{@{}l@{}}\textbf{$\textit{T}_{5}$} \\ \textbf{(ms)} \end{tabular} & \begin{tabular}[c]{@{}l@{}}\textbf{$\textit{T}_{6}$} \\ \textbf{(ms)} \end{tabular} & \begin{tabular}[c]{@{}l@{}}\textbf{$\textit{T}_{7}$} \\ \textbf{(ms)} \end{tabular} & \begin{tabular}[c]{@{}l@{}}\textbf{$\textit{T}_{down}$} \\ \textbf{(ms)} \end{tabular} & \begin{tabular}[c]{@{}l@{}}\textbf{$\textit{T}_{net}$} \\ \textbf{(ms)} \end{tabular} \\ 
\hline
1 & 3.23 & 8.04 & 3.94 & 9.40 & 9.35 & 5.27 & 5.13 & 9.28 & 3.53 & 137.22 \\ 
\hline
2 & 3.22 & 8.05 & 3.95 & 9.38 & 9.38 & 5.27 & 5.13 & 9.28 & 3.53 & 137.22 \\ 
\hline
3 & 3.21 & 8.06 & 3.95 & 9.36 & 9.41 & 5.27 & 5.13 & 9.27 & 3.53 & 137.22 \\ 
\hline
4 & 3.20 & 8.06 & 3.95 & 9.34 & 9.44 & 5.27 & 5.13 & 9.27 & 3.53 & 137.23 \\ 
\hline
5 & 3.19 & 8.07 & 3.96 & 9.33 & 9.47 & 5.27 & 5.13 & 9.26 & 3.52 & 137.23 \\
\hline
\end{tabular}
\arrayrulecolor{black}
\end{table*}

The mean network latency $\textit{$\overline{T}$}_{net}$ is the mean of the network latency of the shortest paths at all time slots, and it can be expressed as
\begin{equation}
\label{eq_30}
\textit{$\overline{T}$}_{net} = \sum_{i=1}^{j} \textit{T}_{net,i}/\textit{j} \text{,}
\end{equation} 
where $\textit{j}$ is the total number of time slots, and $\textit{T}_{net,i}$ is the network latency of the shortest path at the $\textit{i}$th time slot. For example, $\textit{$\overline{T}$}_{net}$ for the first five time slots in Table 3 is equal to 137.22 ms.

\subsection{Average Satellite Transmission Power Calculation}
The transmission power of each satellite is calculated by adding the transmission power required to establish the incoming link and outgoing link for that satellite. For the ingress (or first) and egress (or last) satellites on a shortest path, the incoming link and outgoing link are the uplink and downlink, respectively, and the transmission power for such satellites can be calculated as
\begin{multline}
\label{eq_31}
\textit{P}_{TS,ingress/egress}= \\ 
\textit{P}_R/(\textit{$\eta$}_T\textit{$\eta$}_R\textit{G}_T\textit{G}_R\textit{L}_T\textit{L}_R\textit{L}_A\textit{L}_{PG,up/down}) \\
+ \textit{P}_R/(\textit{$\eta$}_T\textit{$\eta$}_R\textit{G}_T\textit{G}_R\textit{L}_T\textit{L}_R\textit{L}_{PS,out/in}) \text{,}
\end{multline}
where $\textit{L}_{PG,up}$ and $\textit{L}_{PG,down}$ is the free-space path loss for the uplink and downlink, respectively, and $\textit{L}_{PS,in}$ and $\textit{L}_{PS,out}$ is the free-space path loss for the incoming and outgoing ISLs, respectively. The optical links between satellites, i.e., LISLs, and optical links between ground stations and satellites (uplink and downlink) are bidirectional \cite{b36},\cite{b37}. That is why, we are considering the power from the satellite to the ground station for uplink and downlink for modelling transmission power of ingress and egress satellites. For other (or intermediate) satellites on the shortest path, the incoming and outgoing links are ISLs, and the transmission power for such satellites can be calculated as 
\begin{multline}
\label{eq_32}
\textit{P}_{TS,intermediate} = \\
\textit{P}_R/(\textit{$\eta$}_T\textit{$\eta$}_R\textit{G}_T\textit{G}_R\textit{L}_T\textit{L}_R\textit{L}_{PS,in})\\ 
+ \textit{P}_R/(\textit{$\eta$}_T\textit{$\eta$}_R\textit{G}_T\textit{G}_R\textit{L}_T\textit{L}_R\textit{L}_{PS,out}) \text{.}
\end{multline}

The transmission power for each laser link on the shortest paths for the first five time slots for the FSOSN with the Starlink Phase 1 Version 3 constellation at an LISL range of 3,000 km for the Toronto--Sydney inter-continental connection is shown in Table 4. The satellite transmission power and average satellite transmission power for these shortest paths are given in Table 5. In Table 4, $\textit{P}_{up}$ and $\textit{P}_{down}$ represent the transmission power for the uplink and downlink, respectively, and $\textit{P}_{i}$ is the transmission power for the $\textit{i}$th ISL. For example, $\textit{P}_{1}$ is the transmission power for the first ISL on the shortest path at a time slot. In Table 5, the transmission power for the first satellite (or ingress satellite) $\textit{P}_{TS,ingress}$ at the first time slot can be calculated by adding $\textit{P}_{up}$ and $\textit{P}_{1}$ (i.e., 89.34 mW and 198.26 mW, respectively) and is equal to 287.60 mW.

\begin{table*}
\centering
\renewcommand\thetable{4}
\caption{Transmission Power for Uplink/Downlink and Inter-Satellite Links on the Shortest Paths at the First Five Time Slots for the FSOSN with the Starlink Phase 1 Version 3 Constellation at 3,000 km LISL Range for the Toronto--Sydney Inter-Continental Connection.}
\arrayrulecolor{black}
\begin{tabular}{!{\color{black}\vrule}l!{\color{black}\vrule}l!{\color{black}\vrule}l!{\color{black}\vrule}l!{\color{black}\vrule}l!{\color{black}\vrule}l!{\color{black}\vrule}l!{\color{black}\vrule}l!{\color{black}\vrule}l!{\color{black}\vrule}l!{\color{black}\vrule}} 
\hline
\begin{tabular}[c]{@{}l@{}}\textbf{Time} \\ \textbf{Slot} \end{tabular} & \begin{tabular}[c]{@{}l@{}}\textbf{$\textit{P}_{up}$} \\ \textbf{(mW)} \end{tabular} & \begin{tabular}[c]{@{}l@{}}\textbf{$\textit{P}_1$} \\ \textbf{(mW)} \end{tabular} & \begin{tabular}[c]{@{}l@{}}\textbf{$\textit{P}_2$} \\ \textbf{(mW)} \end{tabular} & \begin{tabular}[c]{@{}l@{}}\textbf{$\textit{P}_3$} \\ \textbf{(mW)} \end{tabular} & \begin{tabular}[c]{@{}l@{}}\textbf{$\textit{P}_4$} \\ \textbf{(mW)} \end{tabular} & \begin{tabular}[c]{@{}l@{}}\textbf{$\textit{P}_5$} \\ \textbf{(mW)} \end{tabular} & \begin{tabular}[c]{@{}l@{}}\textbf{$\textit{P}_6$} \\ \textbf{(mW)} \end{tabular} & \begin{tabular}[c]{@{}l@{}}\textbf{$\textit{P}_7$} \\ \textbf{(mW)} \end{tabular} & \begin{tabular}[c]{@{}l@{}}\textbf{$\textit{P}_{down}$} \\ \textbf{(mW)} \end{tabular} \\ 
\hline
1 & 70.42 & 198.26 & 47.67 & 270.82 & 268.28 & 85.14 & 80.80 & 264.18 & 111.49 \\ 
\hline
2 & 69.93 & 198.61 & 47.76 & 269.78 & 269.90 & 85.14 & 80.75 & 263.88 & 111.31 \\ 
\hline
3 & 69.45 & 198.96 & 47.86 & 268.74 & 271.53 & 85.13 & 80.71 & 263.59 & 111.13 \\ 
\hline
4 & 68.99 & 199.31 & 47.95 & 267.71 & 273.17 & 85.12 & 80.67 & 263.29 & 110.97 \\ 
\hline
5 & 68.53 & 199.66 & 48.04 & 266.69 & 274.81 & 85.12 & 80.63 & 262.99 & 110.82 \\
\hline
\end{tabular}
\arrayrulecolor{black}
\end{table*}

\begin{table*}
\centering
\renewcommand\thetable{5}
\caption{Transmission Power for Each Satellite and Average Satellite Transmission Power on the Shortest Paths at the First Five Time Slots for the FSOSN with the Starlink Phase 1 Version 3 Constellation at 3,000 km LISL Range for the Toronto--Sydney Inter-Continental Connection.}
\arrayrulecolor{black}
\begin{tabular}{!{\color{black}\vrule}l!{\color{black}\vrule}l!{\color{black}\vrule}l!{\color{black}\vrule}l!{\color{black}\vrule}l!{\color{black}\vrule}l!{\color{black}\vrule}l!{\color{black}\vrule}l!{\color{black}\vrule}l!{\color{black}\vrule}l!{\color{black}\vrule}} 
\hline
\begin{tabular}[c]{@{}l@{}}\textbf{Time} \\ \textbf{Slot} \end{tabular} & \begin{tabular}[c]{@{}l@{}}\textbf{$\textit{P}_{TS,ingress}$} \\\textbf{(mW)}\end{tabular} & \begin{tabular}[c]{@{}l@{}}\textbf{$\textit{P}_{TS,1}$} \\\textbf{(mW)}\end{tabular} & \begin{tabular}[c]{@{}l@{}}\textbf{$\textit{P}_{TS,2}$} \\\textbf{(mW)}\end{tabular} & \begin{tabular}[c]{@{}l@{}}\textbf{$\textit{P}_{TS,3}$} \\\textbf{(mW)}\end{tabular} & \begin{tabular}[c]{@{}l@{}}\textbf{$\textit{P}_{TS,4}$} \\\textbf{(mW)}\end{tabular} & \begin{tabular}[c]{@{}l@{}}\textbf{$\textit{P}_{TS,5}$} \\\textbf{(mW)}\end{tabular} & \begin{tabular}[c]{@{}l@{}}\textbf{$\textit{P}_{TS,6}$} \\\textbf{(mW)}\end{tabular} & \begin{tabular}[c]{@{}l@{}}\textbf{$\textit{P}_{TS,egress}$} \\\textbf{(mW)}\end{tabular} & \begin{tabular}[c]{@{}l@{}}\textbf{$\textit{P}_{TS,avg}$} \\\textbf{(mW)} \end{tabular} \\
\hline
1 & 268.67 & 245.93 & 318.49 & 539.10 & 353.42 & 165.94 & 344.98 & 375.67 & 326.53 \\ 
\hline
2 & 268.54 & 246.37 & 317.54 & 539.68 & 355.04 & 165.89 & 344.64 & 375.19 & 326.61 \\ 
\hline
3 & 268.41 & 246.81 & 316.60 & 540.27 & 356.66 & 165.84 & 344.30 & 374.72 & 326.70 \\ 
\hline
4 & 268.29 & 247.26 & 315.66 & 540.88 & 358.29 & 165.79 & 343.96 & 374.26 & 326.80 \\ 
\hline
5 & 268.18 & 247.70 & 314.73 & 541.50 & 359.93 & 165.74 & 343.62 & 373.81 & 326.90 \\
\hline
\end{tabular}
\arrayrulecolor{black}
\end{table*}

The average satellite transmission power for satellites on a shortest path at a time slot $\textit{P}_{TS,avg}$ is the average value of the transmission power of all satellites on the shortest path at that time slot, and it can be expressed as
\begin{equation}
\label{eq_33}
\textit{P}_{TS,avg} = (\textit{P}_{TS,ingress} + \sum_{m=1}^{n-2} \textit{P}_{TS,m} + \textit{P}_{TS,egress}) / \textit{n}\text{,}
\end{equation}
where $\textit{P}_{TS,m}$ is the transmission power for the $\textit{m}$th intermediate satellite on the shortest path and there are $\textit{n}$ satellites on the shortest path, and thereby $\textit{n}-2$ intermediate satellites exist in total on the shortest path. The mean of the average satellite transmission power $\textit{$\overline{P}$}_{TS,avg}$, i.e., the mean of the average satellite transmission power of the shortest paths at all time slots, can be expressed as
\begin{equation}
\label{eq_34}
\textit{$\overline{P}$}_{TS,avg} = \sum_{i=1}^{j} \textit{P}_{TS,avg,i}/\textit{j} \text{,}
\end{equation} 
where $\textit{j}$ is the total number of time slots, and $\textit{P}_{TS,avg,i}$ is average satellite transmission power at the $\textit{i}$th time slot. For instance, $\textit{P}_{TS,1}$ is the satellite transmission power for the first intermediate satellite on the shortest path at a time slot in Table 5. The average satellite transmission power $\textit{P}_{TS,avg}$ of the shortest path at the first time slot can be calculated by taking the average value of the transmission powers for the eight satellites on this shortest path, which is equal to 328.89 mW in Table 5. Finally, $\textit{$\overline{P}$}_{TS,avg}$ for the first five time slots in Table 5 is equal to 329.02 mW.

\section{Power versus Latency Tradeoff}
In this work, we assume there is one source ground station in Toronto, and a destination ground station in each of the other cities: Sydney, Istanbul, and London. We then determine the shortest paths for the three inter-continental connections from the source ground station to the destination ground station for the FSOSNs with the two constellations at an LISL range at each time slot. Subsequently, we calculate the network latency and average satellite transmission power for the shortest path at each time slot. In doing so, we are interested in studying the tradeoff between network latency and average satellite transmission power, and we consider several LISL ranges for satellites in each FSOSN as described in Section IV. Finally, we investigate the effect of atmosphere attenuation and turbulence on outage probability of optical uplink/downlink.

\subsection{Simulation Methodology}
For the computation of the average satellite transmission power and network latency of the shortest path for an inter-continental connection, we first simulate the Starlink Phase 1 Version 3 and Kuiper Shell 2 constellations using the well-known satellite constellation simulator Systems Tool Kit (STK) version 12.1 \cite{b38}. We generate a satellite constellation, ground stations, and all possible ISLs and links between satellites and ground stations based on an LISL range within STK, and then we extract the data containing the links between satellites as well as the links between satellites and ground stations and the positions of satellites and ground stations from STK. We process this data in Python to compute the distance for each link at a time slot to find the shortest path between ground stations for an inter-continental connection over the FSOSN at each time slot. To this end, we first discretize the extracted data, as the data obtained from STK is continuous, and for a certain link, it contains the duration of the link for the entire orbital period. We discretize this data by dividing it into time slots and reorganize it so that it shows all links that exist at a particular time slot. The position data for each satellite is also discretized into time slots to match the discretized link data. Next, using the positions of satellites and links between them at each time slot, we compute the distance for a link at each time slot. Finally, we find the shortest path between the source and destination ground stations for an inter-continental connection over the FSOSN using the NetworkX library in Python \cite{b39}.

\subsection{Simulation Setup}
To study the tradeoff between network latency and satellite transmission power in FSOSNs with the Starlink Phase 1 Version 3 and Kuiper Shell 2 constellations, the period of the simulation is taken as 6,000 time slots for both constellations. This encompasses the entire orbital period of a satellite in these FSOSNs, and the duration of a time slot is set to one second. 

To calculate satellite transmission power in the two FSOSNs, we consider the laser link model parameters summarized in Table 6. These parameters are used in existing or practical laser satellite communication systems. We consider On-Off Keying (OOK) as the laser link’s modulation scheme. We assume the wavelength as 1,550 nm for both LISL and laser uplink/downlink, since it is widely used for optical satellite communication due to its better performance. We take both transmitter and receiver optical efficiency as 0.8, as this is the common value of these parameters for laser communication terminals (LCTs) and is used in many simulation studies as well. We assume the data rate as 10 Gbps since this is a practical link data rate for optical satellite communications and the maximum data rate that is offered by Mynaric’s LCT \cite{b40}. The transmitting divergence angle is set to 15 $\mu$rad and the receiver telescope diameter is 80 mm according to Mynaric’s LCT. We assume both transmitter and receiver pointing errors as 1 $\mu$rad, since these values are widely used in LCTs. We use a curve-fitting technique to find the receiver sensitivity of -35.5 dBm for the OOK modulation with a 10 Gbps data rate and $10^{-12}$ BER from \cite{b42}. We assume the ground station altitude as 0.1 km, since this is the general height of tall buildings and towers. We assume clear weather conditions with a thin cirrus cloud concentration of 0.5 cm\textsuperscript{-3} and set liquid water content as $3.128\times10^{-4}$ g/m\textsuperscript{-3} according to \cite{b10}. We take 20 km as the height of the troposphere layer \cite{b44}, since this value is practical in laser uplink/downlink. We consider no special weather condition for all the five inter-continental connection pairs. We consider the $\textit{LM}$ for inter-satellite links as 3 dB and the $\textit{LM}$ for uplink/downlink as 6 dB, since there is more turbulence and attenuation in uplink/downlink. This yields a received power $\textit{P}_R$ of -32.5 dBm for ISLs and -29.5 dBm for uplink/downlink according to (23). To calculate the OP for uplink/downlink, we assume the given SNR threshold as 7 dB, and elevation angle as 30 degrees. As mentioned earlier, we consider three cloud scenarios, each characterized by different parameters: thin cirrus clouds, cirrus clouds, and cumulus cloud concentrations.

\begin{table}
\centering
\renewcommand\thetable{6}
\caption{Simulation Parameters.}
\setlength{\tabcolsep}{4pt}
\arrayrulecolor{black}
\begin{tabular}{llll} 
\hline
\textbf{Parameter} & \textbf{Symbol} & \textbf{Units} & \textbf{Value} \\ 
\hline
Laser wavelength \cite{b40} & $\textit{$\lambda$}$ & nm & 1550 \\
Transmitter optical efficiency \cite{b41} & $\textit{$\eta$}_T$ & ~ & 0.8 \\
Receiver optical efficiency \cite{b41} & $\textit{$\eta$}_R$ & ~ & 0.8 \\
Data rate \cite{b40} & $\textit{R}_{data}$ & Gbps & 10 \\
Receiver telescope diameter \cite{b40} & $\textit{D}_R$ & mm & 80 \\
Transmitter pointing error \cite{b41} & $\textit{$\theta$}_T$ & $\mu$rad & 1 \\
Receiver pointing error \cite{b41} & $\textit{$\theta$}_R$ & $\mu$rad & 1 \\
\begin{tabular}[c]{@{}l@{}}Full transmitting divergence \\ angle \cite{b42}\end{tabular} & $\textit{$\Theta$}_T$ & $\mu$rad & 1.5 \\
Receiver sensitivity \cite{b42} &$ \textit{P}_{req}$ & dBm & -35.5 \\
Bit error rate \cite{b42} & ~ & ~ & 10\textsuperscript{-12} \\
Ground station attitude & $\textit{h}_E$ & km & 0.1 \\
Thin cirrus cloud concentration \cite{b10} & $\textit{L}_{W}$ & cm\textsuperscript{-3} & 0.5 \\
Thin cirrus liquid water content \cite{b10} & \textit{N} & g/m\textsuperscript{-3} & 3.128×10\textsuperscript{-4} \\
Cirrus cloud concentration \cite{b10} & $\textit{L}_{W}$ & cm\textsuperscript{-3} & 0.0255 \\
Cirrus liquid water content \cite{b10} & \textit{N} & g/m\textsuperscript{-3} & 0.06405 \\
Cumulus cloud concentration \cite{b10} & $\textit{L}_{W}$ & cm\textsuperscript{-3} & 250 \\
Cumulus liquid water content \cite{b10} & \textit{N} & g/m\textsuperscript{-3} & 1.0 \\
SNR threshold \cite{b43} &  $\textit{$\gamma$}_{th}$  & dB  & 7 \\
Elevation angle  &  $\textit{$\theta$}_{E}$  & degree  & 30 \\
Partial size coefficient \cite{b43} & \textit{$\Phi$} & ~ & 1.6 \\
Troposphere layer height \cite{b44} & $\textit{h}_A$ & km & 20 \\
Atmosphere layer height \cite{b2} & $\textit{h}_a$ & km & 80 \\
\begin{tabular}[c]{@{}l@{}}Location of source \\ ground station\end{tabular} & \textit{~} & ~ & Toronto \\
\begin{tabular}[c]{@{}l@{}}Location of destination \\ ground station\end{tabular} & \textit{~} & ~ & \begin{tabular}[c]{@{}l@{}@{}}\{Sydney, \\ Istanbul, \\ London\} \end{tabular} \\
Node delay & $\textit{T}_{node}$ & ms & 10 \\
LISL range for Kuiper Shell 2 & \textit{~} & km & \begin{tabular}[c]{@{}l@{}@{}}\{1515, 2000, \\ 3000, 4000, \\ 5339\} \end{tabular} \\
\begin{tabular}[c]{@{}l@{}}LISL range for Starlink Phase 1 \\ Version 3 \end{tabular}& \textit{~} & km & \begin{tabular}[c]{@{}l@{}@{}}\{1575, 1731, \\ 2000, 3000, \\ 4000, 5016\} \end{tabular} \\
Speed of light & $\textit{c}_s$ & m/s & 299792458 \\
Duration of a time slot & ~ & s & 1 \\
Total number of time slots & \textit{j} & & 6000 \\
\hline
\end{tabular}
\arrayrulecolor{black}
\end{table}

\subsection{$\textit{$\overline{T}$}_{net}$ versus $\textit{$\overline{P}$}_{TS,avg}$ for Different LISL Ranges}
To show the relationship and tradeoff between network latency and satellite transmission power, we superimpose the two curves representing $\textit{$\overline{T}$}_{net}$ and $\textit{$\overline{P}$}_{TS,avg}$ for different LISL ranges on a unified plot. For different inter-continental connections, the maximum value of $\textit{$\overline{T}$}_{net}$ (with the minimum corresponding to the $\textit{$\overline{P}$}_{TS,avg}$ case) is realized when the LISL range is at its minimum feasible value for each FSOSN. Conversely, the minimum value of $\textit{$\overline{T}$}_{net}$ (with the maximum corresponding to the $\textit{$\overline{P}$}_{TS,avg}$ case) is observed at the maximum feasible LISL range. Within the plot, the curve of $\textit{$\overline{T}$}_{net}$ gradually descends over LISL ranges, whereas the curve of $\textit{$\overline{P}$}_{TS,avg}$ rises with the increasing LISL range. The point of the intersection between these two curves signifies the juncture of relatively optimal performance between network latency and satellite transmission power can be reached at that LISL range for a given inter-continental connection. Any deviation from this identified LISL range could lead to higher energy consumption or worse latency performance, underscoring the critical importance of maintaining this delicate balance in a FSOSN.

Figs. 5 and 6 show the relationship between $\textit{$\overline{T}$}_{net}$ and $\textit{$\overline{P}$}_{TS,avg}$ for different LISL ranges for the two FSOSNs for the Toronto--Sydney inter-continental connection. For the FSOSN with the Starlink Phase 1 Version 3 constellation, the curves of $\textit{$\overline{T}$}_{net}$ and $\textit{$\overline{P}$}_{TS,avg}$ intersect when the LISL range is approximately 2,900 km, which indicates that $\textit{$\overline{T}$}_{net}$ and $\textit{$\overline{P}$}_{TS,avg}$ are balanced when the LISL range is 2,900 km for satellites in this FSOSN for this inter-continental connection, and $\textit{$\overline{T}$}_{net}$ and $\textit{$\overline{P}$}_{TS,avg}$ are around 135 ms and 380 mW, respectively, at this intersection. For the FSOSN with the Kuiper Shell 2 constellation, $\textit{$\overline{T}$}_{net}$ and $\textit{$\overline{P}$}_{TS,avg}$ intersect when the LISL range is around 3,800 km for this inter-continental connection, and these two parameters are approximately 120 ms and 700 mW, respectively. 

In Figs. 7 and 8, we can see the tradeoff between $\textit{$\overline{T}$}_{net}$ and $\textit{$\overline{P}$}_{TS,avg}$ at different LISL ranges in the two FSOSNs for the Toronto--Istanbul inter-continental connection, and Figs. 9 and 10 show this tradeoff in the two FSOSNs for the Toronto--London inter-continental connection. For the Toronto--Istanbul inter-continental connection, $\textit{$\overline{T}$}_{net}$ and $\textit{$\overline{P}$}_{TS,avg}$ intersect when the LISL range is about 2,600 km for the FSOSN with the Starlink Phase 1 Version 3 constellation and 2,900 km for the FSOSN with the Kuiper Shell 2 constellation. For the Toronto to London inter-continental connection, these two parameters intersect when the LISL ranges are 3,400 km and 3,000 km for the FSOSNs with the Starlink Phase 1 Version 3 and Kuiper Shell 2 constellations, respectively. For both FSOSNs in all inter-continental connections, as the LISL range increases, higher transmission power is needed for a satellite but lower network latency is obtained. For $\textit{$\overline{T}$}_{net}$ curve in Fig. 9, it changes sightly between 3000 km to 4000 km LISL range, which is different from $\textit{$\overline{T}$}_{net}$ at the same LISL range for Kuiper Shell 2 constellation as shown in Fig. 10. This is because that the distribution of the satellites for Starlink Phase 1 Version 3 constellation is denser at high latitude comparing to Kuiper Shell 2 constellation, in this way for short inter-continental connections like Toronto to London $\textit{$\overline{T}$}_{net}$ can change unobtrusively at certain LISL ranges.

\begin{figure}[htbp]
\centerline{\includegraphics[width=0.4788\textwidth]{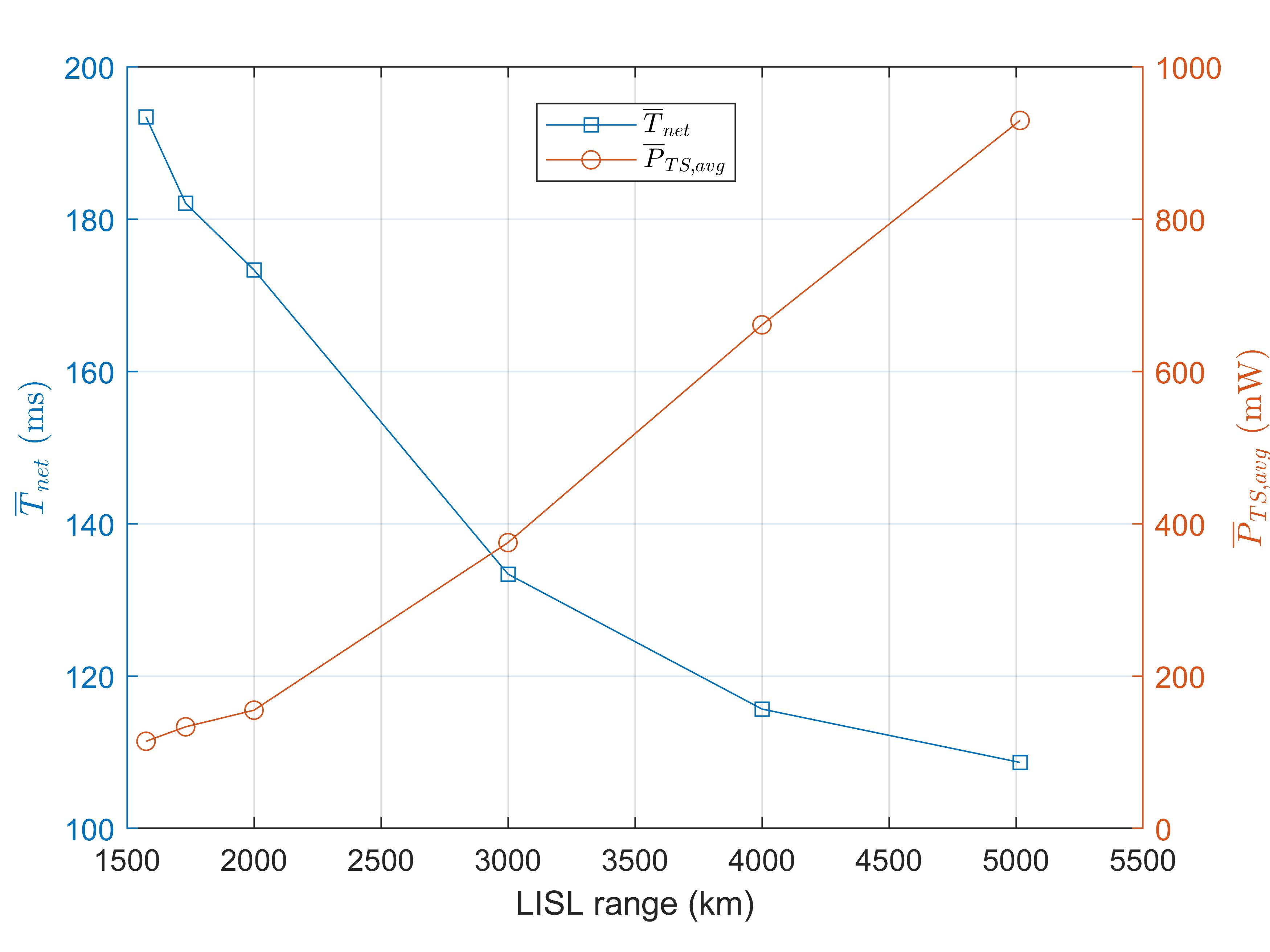}}
\caption{$\textit{$\overline{T}$}_{net}$ vs. $\textit{$\overline{P}$}_{TS,avg}$ at different LISL ranges for the FSOSN with the Starlink Phase 1 Version 3 constellation for the Toronto--Sydney inter-continental connection.}
\label{fig}
\end{figure}
\begin{figure}[htbp]
\centerline{\includegraphics[width=0.4788\textwidth]{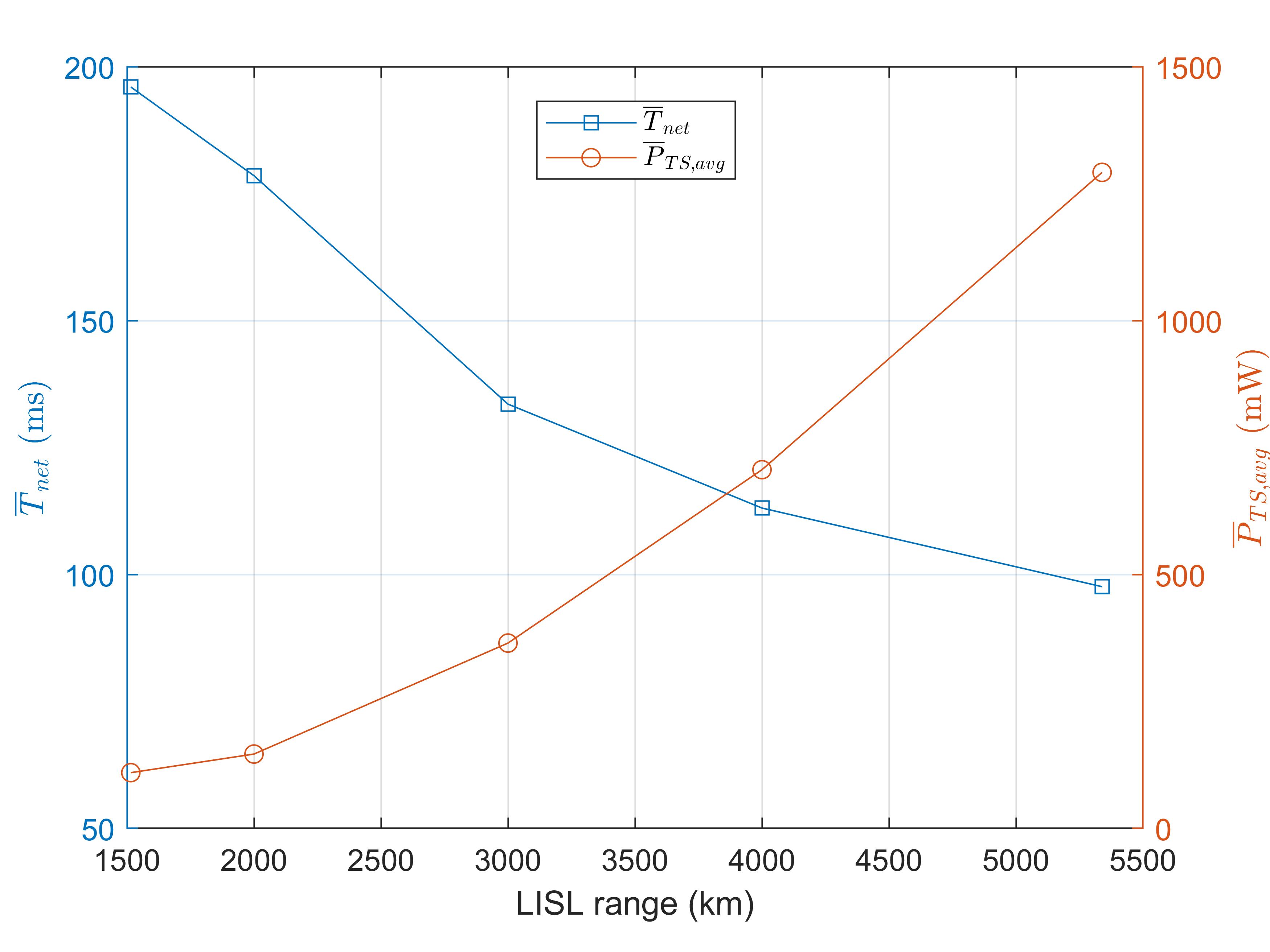}}
\caption{$\textit{$\overline{T}$}_{net}$ vs. $\textit{$\overline{P}$}_{TS,avg}$ at different LISL ranges for the FSOSN with the Kuiper Shell 2 constellation for the Toronto--Sydney inter-continental connection.}
\label{fig}
\end{figure}
\begin{figure}[htbp]
\centerline{\includegraphics[width=0.4788\textwidth]{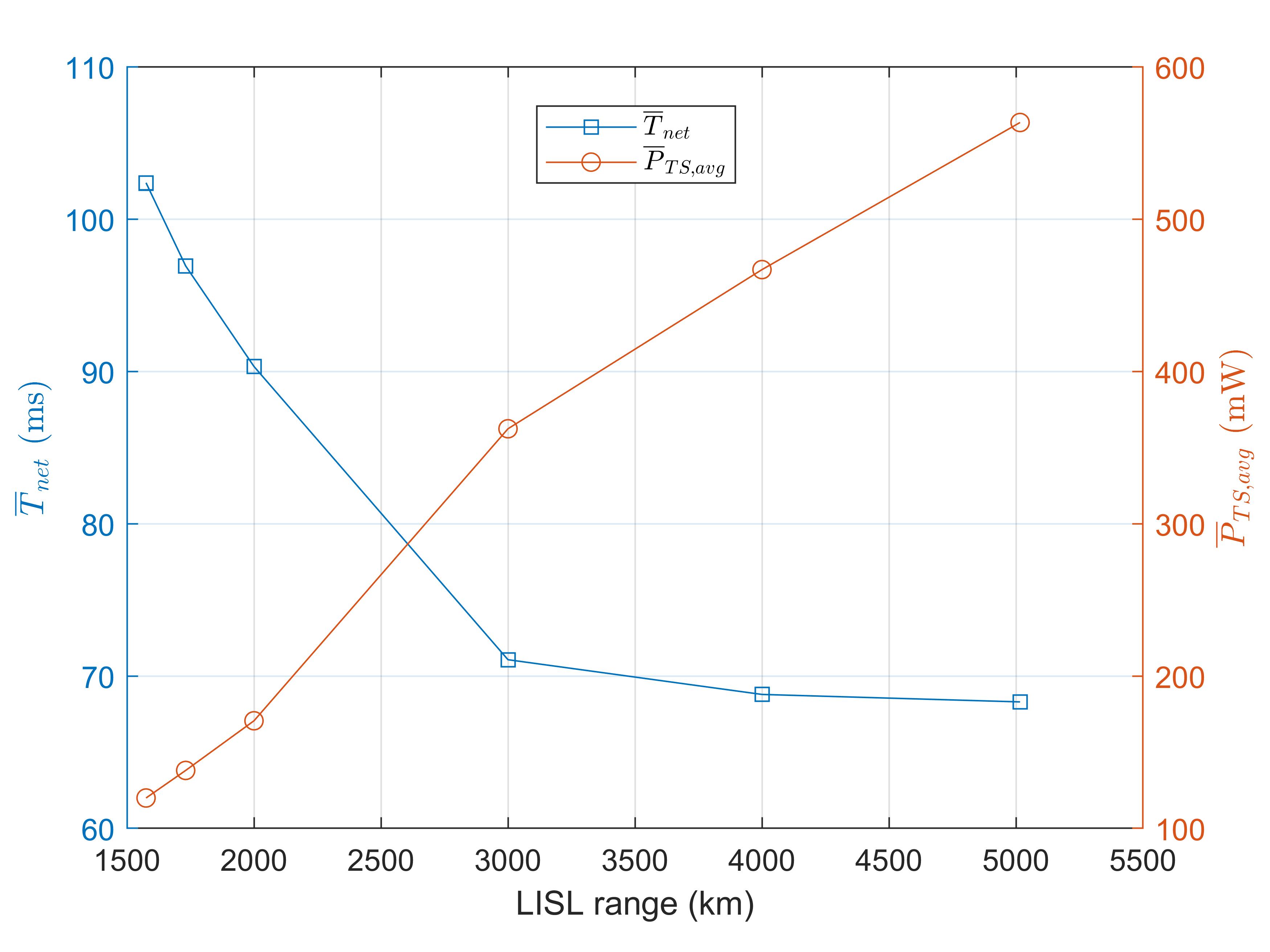}}
\caption{$\textit{$\overline{T}$}_{net}$ vs. $\textit{$\overline{P}$}_{TS,avg}$ at different LISL ranges for the FSOSN with the Starlink Phase 1 Version 3 constellation for the Toronto--Istanbul inter-continental connection.}
\label{fig}
\end{figure}
\begin{figure}[htbp]
\centerline{\includegraphics[width=0.4788\textwidth]{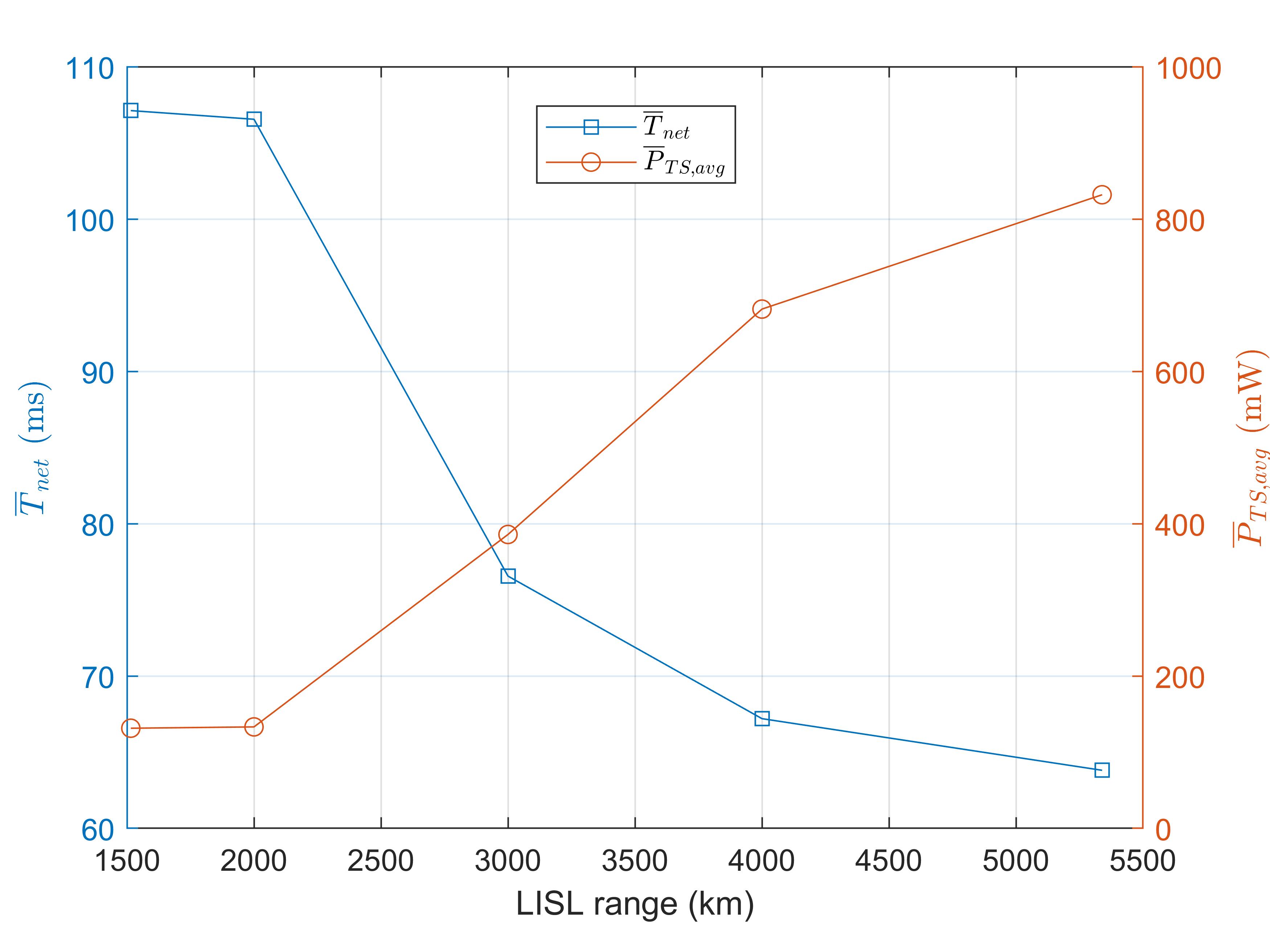}}
\caption{$\textit{$\overline{T}$}_{net}$ vs. $\textit{$\overline{P}$}_{TS,avg}$ at different LISL ranges for the FSOSN with the Kuiper Shell 2 constellation for Toronto--Istanbul inter-continental connection.}
\label{fig}
\end{figure}
\begin{figure}[htbp]
\centerline{\includegraphics[width=0.4788\textwidth]{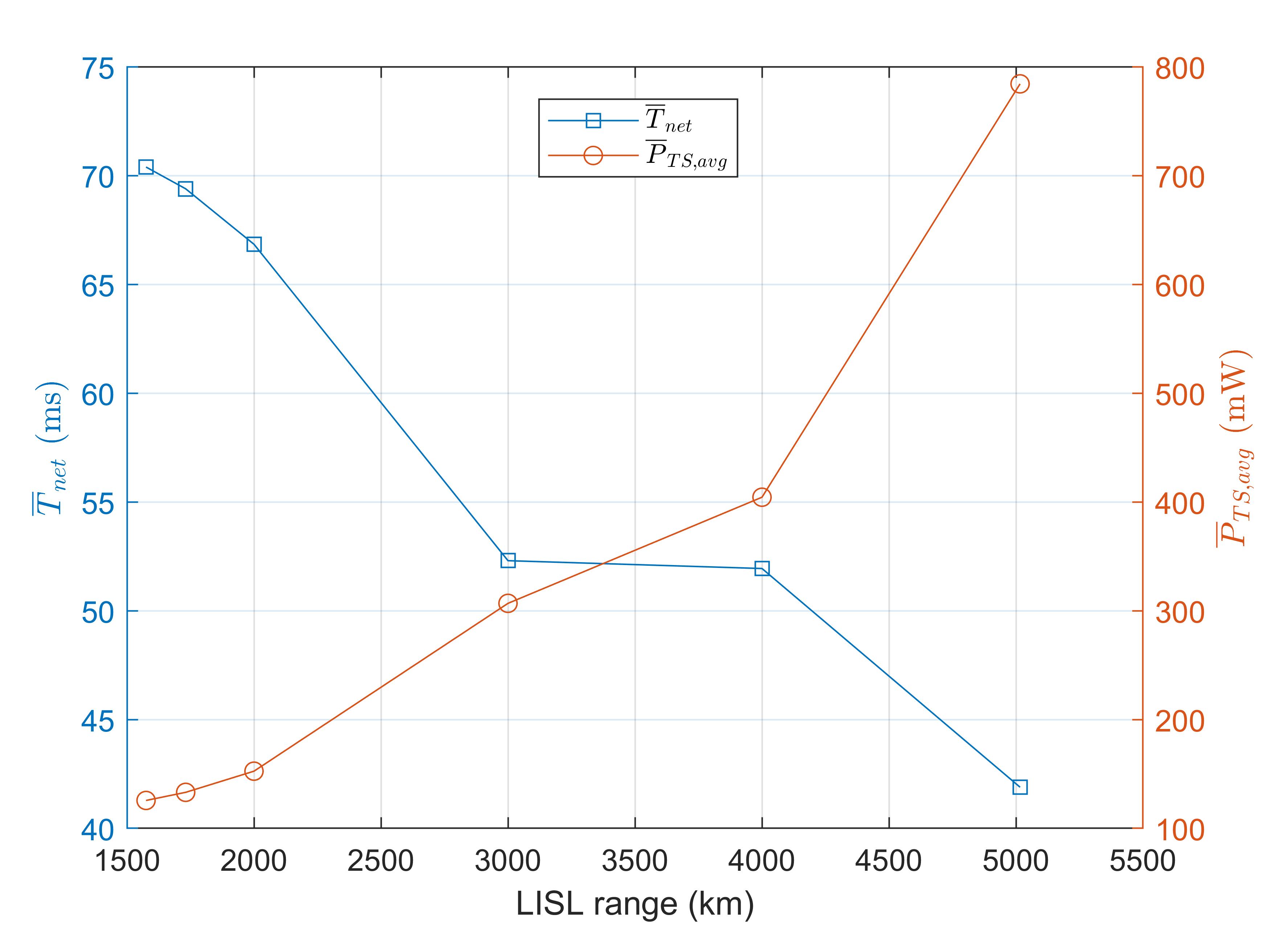}}
\caption{$\textit{$\overline{T}$}_{net}$ vs. $\textit{$\overline{P}$}_{TS,avg}$ at different LISL ranges for the FSOSN with the Starlink Phase 1 Version 3 constellation for the Toronto--London inter-continental connection.}
\label{fig}
\end{figure}
\begin{figure}[htbp]
\centerline{\includegraphics[width=0.4788\textwidth]{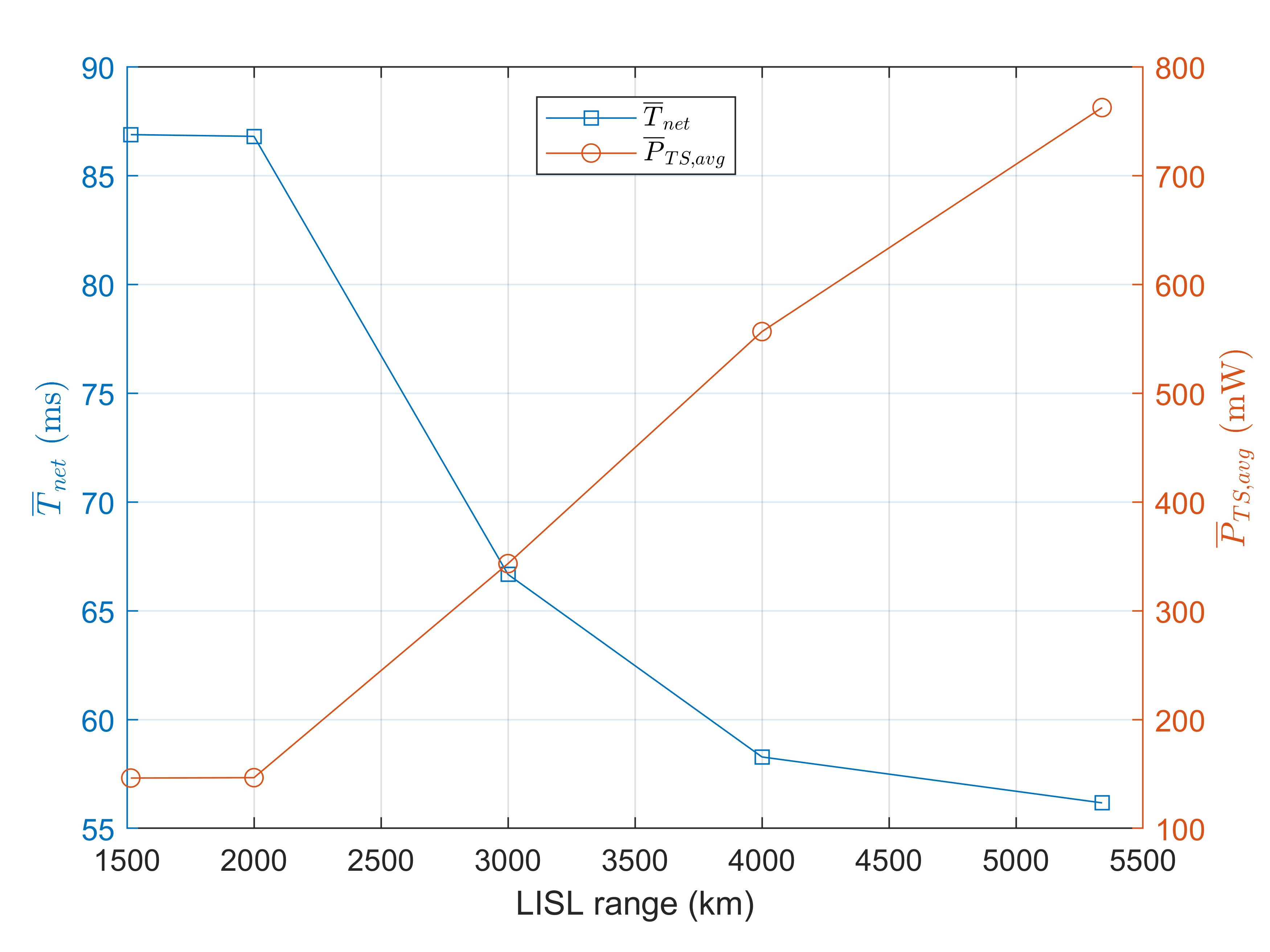}}
\caption{$\textit{$\overline{T}$}_{net}$ vs. $\textit{$\overline{P}$}_{TS,avg}$ at different LISL ranges for the FSOSN with the Kuiper Shell 2 constellation for the Toronto--London inter-continental connection.}
\label{fig}
\end{figure}

\subsection{Outage Probability Analysis}
Figs. 11 and 12 show the relationship between OP and SNR for optical uplink and downlink, respectively. Based on Table 6, we consider three different weather conditions as thin cirrus, cirrus and cumulus clouds. We assume the average SNR as 30 dB, the SNR threshold as 7 dB, the elevation angle of the ground station as 30 degrees and employing Starlink Phase 1 Version 3 constellation, we plot the two figures. As depicted in Figs. 11 and 12, the OP decreases as the SNR increases under cirrus and thin cirrus cloud conditions. Notably, the thin cirrus cloud condition exhibits better performance compared to the cirrus cloud condition in terms of OP. The OP under cumulus cloud condition is 1.0, which indicates an unapproachable optical communication. Comparing the curves of the two figures under same cloud condition, we find that the OP performance for uplink is better than downlink since less atmosphere attenuation ocurrs through uplink. For upink communication, SNR should reach 45 dB to obtain $10^{-8}$ OP performance under thin cirrus cloud condition; while for downlink communication, the requirement of SNR to achieve the same OP performance is 57 dB for thin cirrus cloud condition.
\begin{figure}[htbp]
\centerline{\includegraphics[width=0.4788\textwidth]{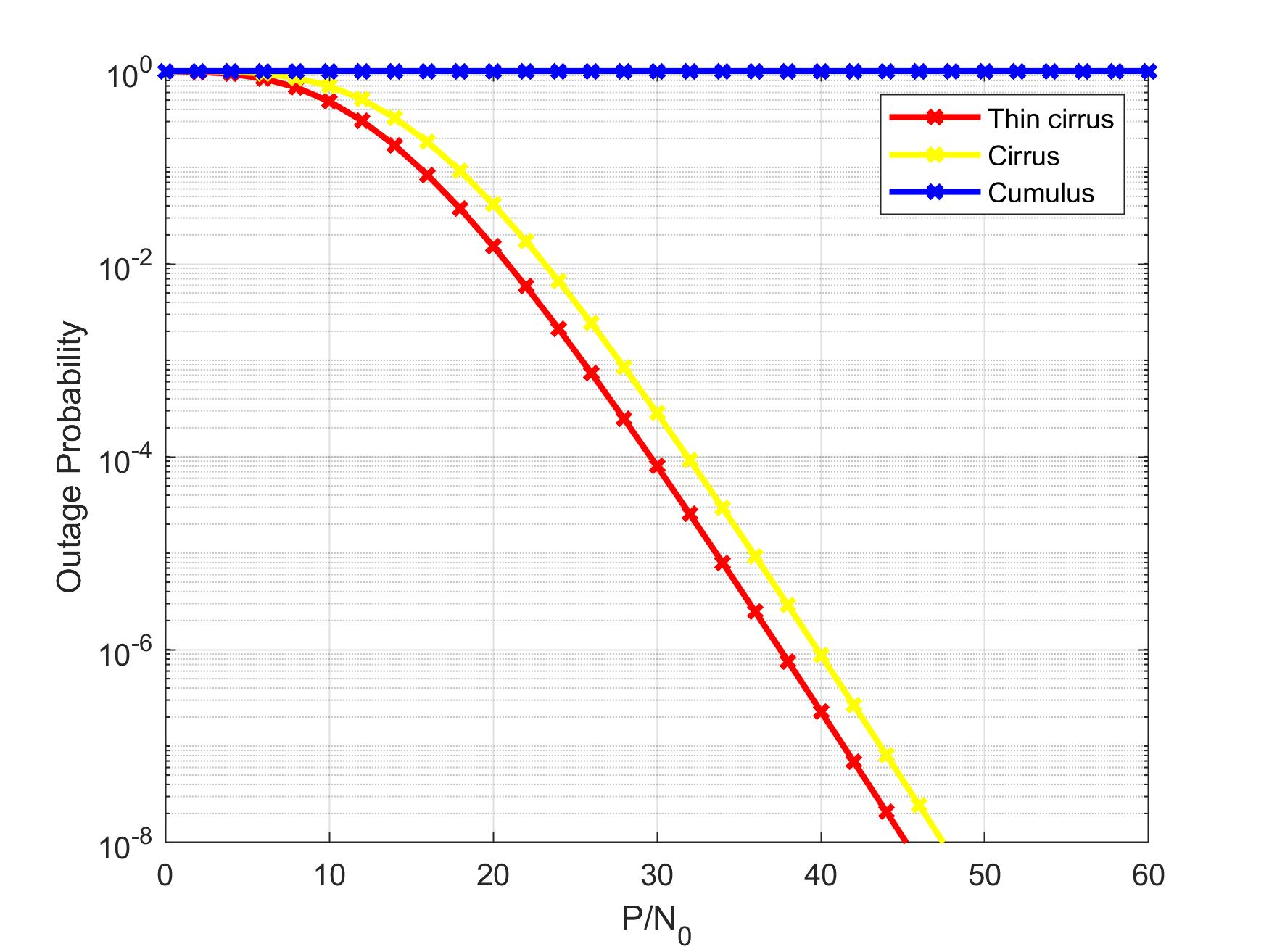}}
\caption{Outage probability vs. $\textit{P}$/$\textit{N}_0$ under different weather conditions for optical uplink communications.}
\label{fig}
\end{figure}\begin{figure}[htbp]
\centerline{\includegraphics[width=0.4788\textwidth]{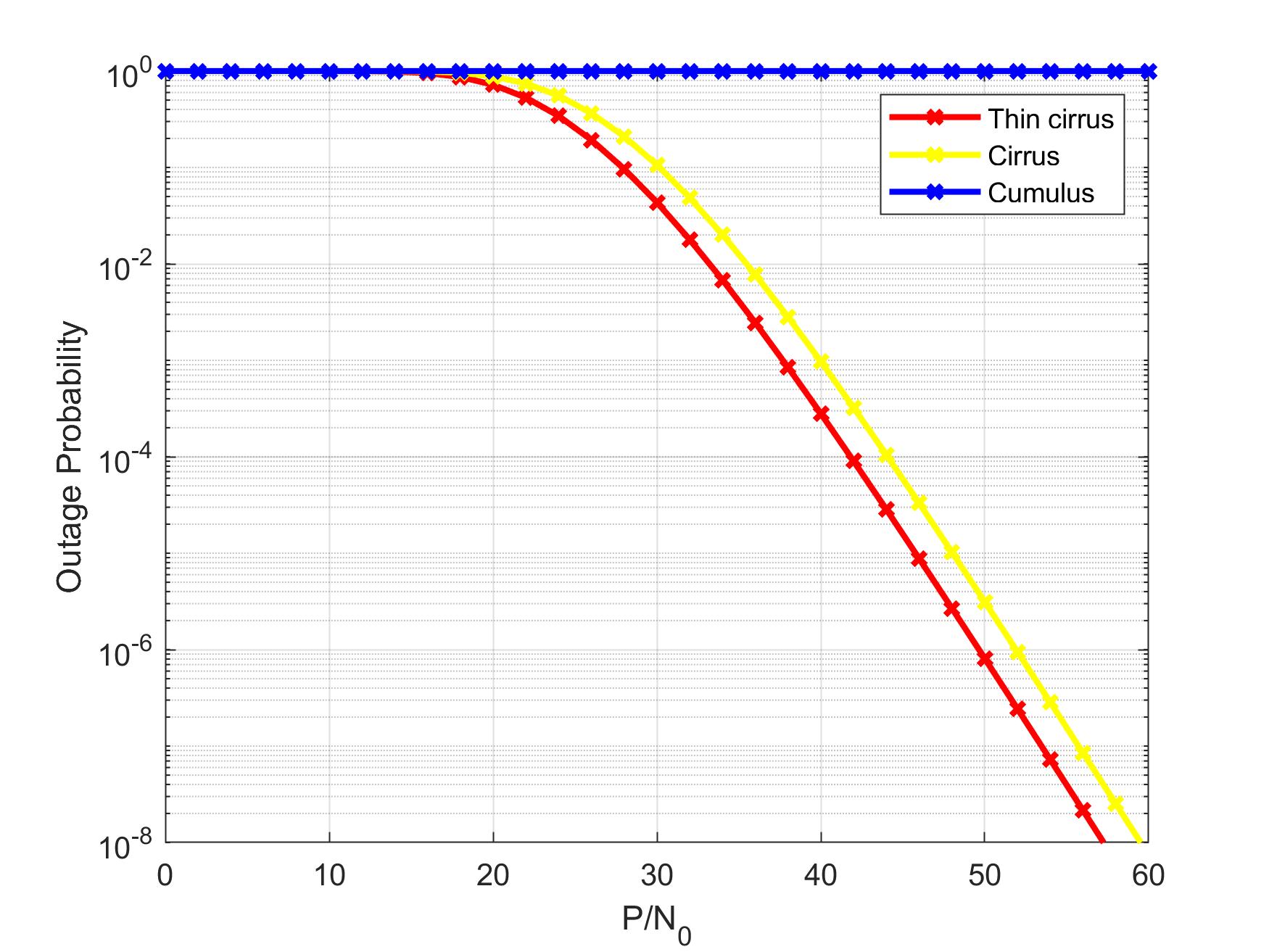}}
\caption{Outage probability vs. $\textit{P}$/$\textit{N}_0$ under different weather conditions for optical downlink communications.}
\label{fig}
\end{figure}

\subsection{Insights}
Here we provide important practical insights for FSOSNs based on the analysis in this work.

$\bullet$ For LISL and laser uplink/downlink, the network latency for the shortest path for an inter-continental connection between two ground stations and the average satellite transmission power for the shortest path have an inverse relationship. As the LISL range increases, the links on the shortest path become longer, the propagation distance of the laser links increases, and the transmission power for the laser links increases along with the average satellite transmission power. At the same time, there are fewer direct links and nodes on the shortest path, and accordingly, the total propagation delay of the path decreases, the total node delay of the path decreases, and the network latency of the path decreases. Therefore, according to this reverse relationship, in order to reach less network latency to achieve better network performance, satellites needed to be designed for high energy payload to sustain higher transmission power consumption. This can be accomplished by implementing larger capacity battery, applying dynamic ISLs instead of establishing the links throughout the entire working period.

$\bullet$ Based on the results of the tradeoff analysis between $\textit{$\overline{T}$}_{net}$ and $\textit{$\overline{P}$}_{TS,avg}$ for different LISL ranges for the two FSOSNs for different inter-continental connections, we find that there is no clear relationship between the LISL range when $\textit{$\overline{T}$}_{net}$ and $\textit{$\overline{P}$}_{TS,avg}$ intersect and the length of the inter-continental connection (i.e., the end-to-end distance between the source and destination ground stations for an inter-continental connection over an FSOSN) within the same FSOSN. A shorter inter-continental connection may have a longer LISL range when $\textit{$\overline{T}$}_{net}$ and $\textit{$\overline{P}$}_{TS,avg}$ intersect. As we can observe that there are different LISL ranges where $\textit{$\overline{T}$}_{net}$ and $\textit{$\overline{P}$}_{TS,avg}$ reach a balance for different inter-continental connections over an FSOSN, all inter-continental connections over an FSOSN must be considered simultaneously while determining an appropriate LISL range for the FSOSN to balance the average satellite transmission power and network latency.

$\bullet$ In laser uplink/downlink communications, the performance of the outage probability is inevitably influenced by cloud conditions. In the case of adverse cloud conditions, such as cumulus clouds, the connectivity of the laser link becomes unavailable, irrespective of the SNR. Under these circumstances, it becomes necessary to transmit through alternative laser links between the ground station and satellites, circumventing the area affected by poor cloud conditions, even though this may cost more network latency and higher satellite transmission power. For laser downlink communication, higher SNR is required to sustain an acceptable OP comparing to uplink communication, and more satellite transmission power will be consumed.

\section{Conclusion}
In this work, we investigated the tradeoff between satellite transmission power and network latency in FSOSNs resulting from the Starlink Phase 1 Version 3 and Kuiper Shell 2 constellations. We used appropriate system models for laser link transmission power and latency and studied $\textit{$\overline{T}$}_{net}$ and $\textit{$\overline{P}$}_{TS,avg}$ in the two FSOSNs for different inter-continental connections and different LISL ranges. We further studied the effect of atmosphere attenuation and turbulence on outage probability of laser uplink/downlink.

We found that as the LISL range increases, $\textit{$\overline{P}$}_{TS,avg}$ increases while $\textit{$\overline{T}$}_{net}$ decreases for both FSOSNs in all inter-continental connections. The results show that for the inter-continental connection from Toronto to Sydney in the FSOSN based on the Starlink Phase 1 Version 3 constellation, $\textit{$\overline{T}$}_{net}$ and $\textit{$\overline{P}$}_{TS,avg}$ intersect when the LISL range is approximately 2,900 km, with these parameters being approximately 135 ms and 380 mW, respectively. For the FSOSN based on the Kuiper Shell 2 constellation in this inter-continental connection, this LISL range is around 3,800 km, and the two parameters are approximately 120 ms and 700 mW, respectively. For the Toronto--Istanbul inter-continental connection, these parameters intersect when LISL ranges are about 2,600 km for the FSOSN based on the Starlink Phase 1 Version 3 constellation and 2,900 km for the FSOSN based on the Kuiper Shell 2 constellation. For the Toronto to London inter-continental connection, $\textit{$\overline{T}$}_{net}$ and $\textit{$\overline{P}$}_{TS,avg}$ intersect when LISL ranges are 3,400 km and 3,000 km for the FSOSNs for the Starlink Phase 1 Version 3 and Kuiper Shell 2 constellation, respectively. The OP decreases as the SNR increases under cirrus and thin cirrus cloud conditions, while the thin cirrus cloud condition performs better compared to the cirrus cloud condition. For cumulus cloud condition the connectivity of the laser link is unreachable. Furthermore, the OP performance for uplink is better than downlink since less atmosphere attenuation ocurrs through uplink.

In a future work, we will investigate the problem of determining an appropriate LISL range for balancing network latency and average satellite transmission power in an FSOSN. This will require the problem to be formulated as a multi-objective mathematical programming problem. We will further examine the problem of determining an appropriate LISL range for an FSOSN for balancing average satellite transmission power and network latency over multiple inter-continental connections (multiple source-destination pairs) simultaneously for that FSOSN. A LEO satellite constellation at a higher altitude requires fewer satellites to achieve the same coverage, which can affect the network latency and average satellite transmission power, and we will investigate the effect of satellite constellation altitude on the tradeoff between network latency and satellite transmission power. The power consumption for optical amplification and signal regeneration is also important in terms of the total power consumption, we will examine the tradeoff between network latency and the satellite total power consumption in future. In this work, we haven't considered the mobility of the satellites and its impact on the tradeoff between network latency and average satellite transmission power, and we will examine this important factor in future. In this work, we don't consider the effect of different satellite constellation parameters, such as number of orbital planes, number of satellites per plane, inclination angle, etc., on the tradeoff between network latency and average satellite transmission power, and we will examine the effect of these parameters in future.


\appendix

\section{Dijkstra's algorithm}
The Dijkstra's algorithm we discissed in section III are also described in the pseudocode in Algorithm 1, where \textit{Y} is the set for the satellites and ground stations; \textit{dist}($\textit{x}_i$) is the distance for a node $\textit{x}_i$ (i.e., a satellite or $\textit{g}_d$) to the source ground station $\textit{g}_{s}$; and $\textit{d}_{q,p_i}$ is the distance of the laser link between nodes \textit{q} and $\textit{p}_i$.

\begin{algorithm}
\caption{Dijkstra's algorithm to find a shortest path over an FSOSN.}\label{alg:cap}
\textbf{Inputs}: \\
$\bullet$ Adjacency matrix containing propagation distances of laser links between satellites and between satellites and ground stations\\
$\bullet$ \textit{Y}, i.e., the set for the satellites and ground stations\\
\textbf{Output}: \\
$\bullet$ Shortest path from $\textit{g}_s$ to $\textit{g}_d$\\
\textbf{BEGIN}
\begin{algorithmic}[1]
\State \textit{S} = \{\}
\State \textit{X} = \textit{Y}
\State \textit{D}($\textit{x}_i$) = \textit{$\infty$} for each node $\textit{x}_i$ in \textit{X}
\State \textit{D}($\textit{g}_s$) = 0
\State move $\textit{g}_s$ from \textit{X} to \textit{S} 
\State \textbf{While} $\textit{g}_d$ $\notin$ \textit{S}
\State ~~~~pick \textit{q} in \textit{X} that is closest to $\textit{g}_{s}$
\State ~~~~move \textit{q} from \textit{X} to \textit{S}
\State ~~~~update \textit{D}(\textit{q}) = \textit{dist}(\textit{q}) 
\State ~~~~\textbf{For} each $\textit{p}_i$ adjacent to \textit{q} in \textit{X}
\State ~~~~~~~~\textbf{If} \textit{dist}($\textit{p}_i$) \textgreater{} \textit{dist}(\textit{q}) + $\textit{d}_{q,p_i}$ 
\State ~~~~~~~~~~~~\textit{dist}($\textit{p}_i$) = \textit{dist}(\textit{q}) + $\textit{d}_{q,p_i}$
\State ~~~~~~~~\textbf{End If}
\State ~~~~\textbf{End For}
\State \textbf{End While}
\State Output \textit{S}, i.e., the set containing the shortest path from $\textit{g}_s$ to $\textit{g}_d$
\end{algorithmic}
\textbf{END}
\end{algorithm}

\begin{IEEEbiography}[{\includegraphics[width=1in,height=1.25in,clip,keepaspectratio]{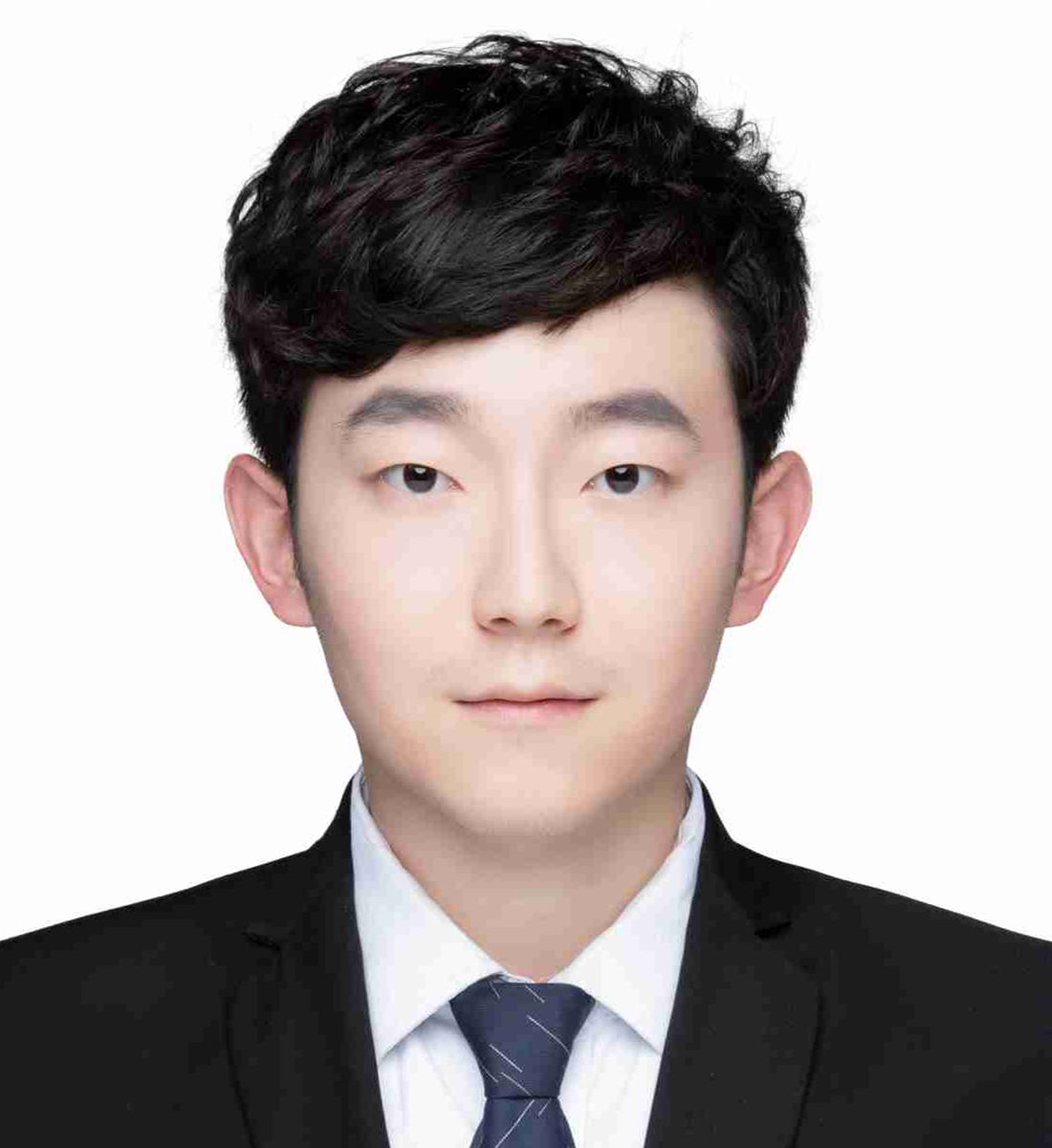}}]{Jintao Liang}
\:(Member, IEEE) received his Bachelor of Engineering degree in Automation Engineering from Shandong University (China) in 2019 and his M.A.Sc degree in Electrical and Computer Engineering from Carleton University (Canada) in 2022. \par
He is currently a Ph.D. student in the Department of Systems and Computer Engineering in Carleton University (Canada) since 2023. \par
Mr. Liang's research interest includes free-space optical satellite networks.\par
\end{IEEEbiography}

\begin{IEEEbiography}[{\includegraphics[width=1in,height=1.25in,clip,keepaspectratio]{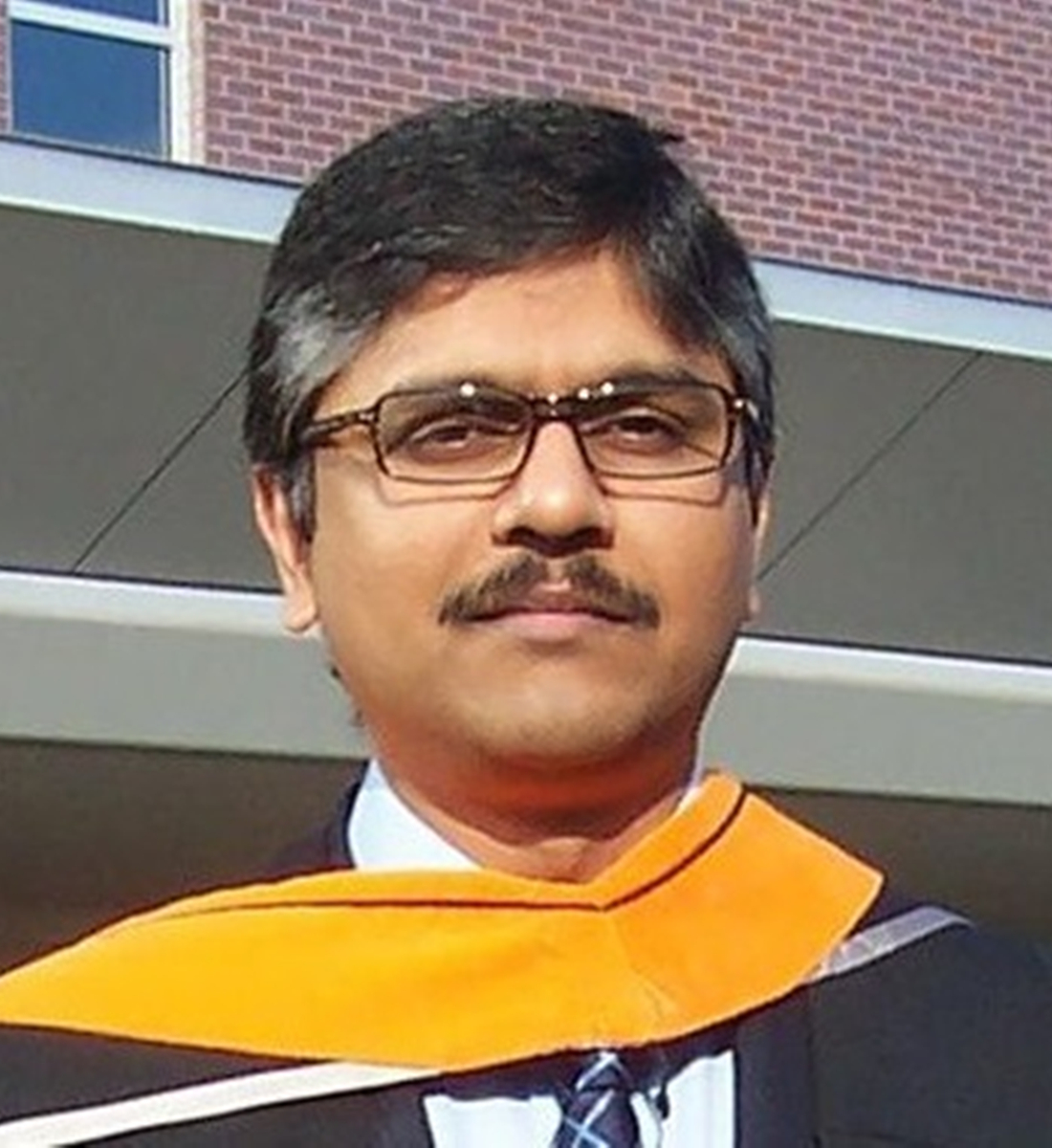}}]{Aizaz U. Chaudhry}
\:(Senior Member, IEEE) received his B.Sc. degree in Electrical Engineering from the University of Engineering and Technology Lahore (Pakistan) in 1999, and his M.A.Sc. and Ph.D. degrees in Electrical and Computer Engineering from Carleton University (Canada) in 2010 and 2015, respectively.\par
He is currently a Senior Research Associate with the Department of Systems and Computer Engineering at Carleton University. Previously, he worked as an NSERC Postdoctoral Research Fellow at Communications Research Centre Canada, Ottawa. His research work has been published in refereed venues, and has received several citations. He has authored and co-authored more than thirty-five publications. His research interests include the application of machine learning and optimization in wireless networks.\par
Dr. Chaudhry is a licensed Professional Engineer in the Province of Ontario, a Senior Member of IEEE, and a Member of IEEE ComSoc’s Technical Committee on Satellite and Space Communications. He serves as a technical reviewer for conferences and journals on a regular basis. He has also served as a TPC Member for conferences, such as IEEE ICC 2021 Workshop 6GSatComNet, IEEE ICC 2022 Workshop 6GSatComNet, IEEE VTC 2022-Spring, IEEE VTC 2022-Fall, the twenty-first international conference on networks (ICN 2022), eleventh advanced satellite multimedia systems conference (ASMS 2022), seventeenth signal processing for space communications workshop (SPSC 2022), and the first international symposium on satellite communication systems and services (SCSS 2022).\par
\end{IEEEbiography}

\begin{IEEEbiography}[{\includegraphics[width=1in,height=1.25in,clip,keepaspectratio]{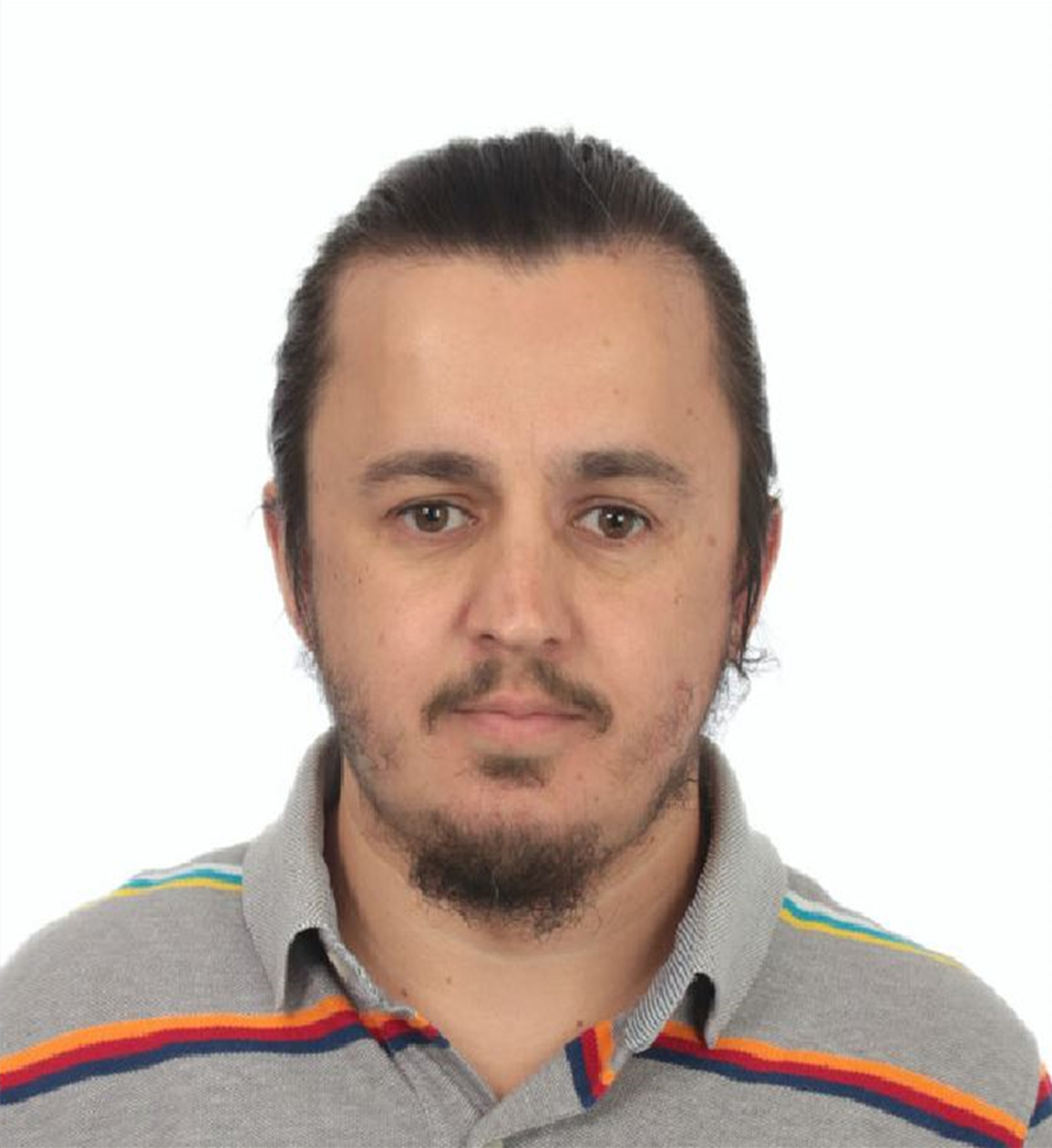}}]{Eylem Erdogan}
\:(Senior Member, IEEE) received B.Sc. and M.Sc. degree from Işık University, Istanbul, Turkey and the Ph.D. degree from Kadir Has University, Istanbul, Turkey in 2014 all in electronics engineering. \par
He is currently an Associate Professor in the Department of Electrical and Electronics Engineering, Istanbul Medeniyet University. He was a Post-Doctoral Fellow in Electrical Engineering department, Lakehead University, Thunder Bay, ON, Canada from March 2015 to September 2016 and a visiting professor in Carleton University, Ottawa, Canada during summer 2019. He has authored or coauthored more than 40+ papers in peer reviewed SCI/SCI-E journals and international conferences. \par
His research interests are in the broad areas of wireless communications, including signal processing for wireless communications, the performance analysis of cooperative relaying in cognitive radio networks, unmanned aerial vehicle communications and networks and free space optical communications. He is a Senior Member of IEEE. \par
\end{IEEEbiography}

\begin{IEEEbiography}[{\includegraphics[width=1in,height=1.25in,clip,keepaspectratio]{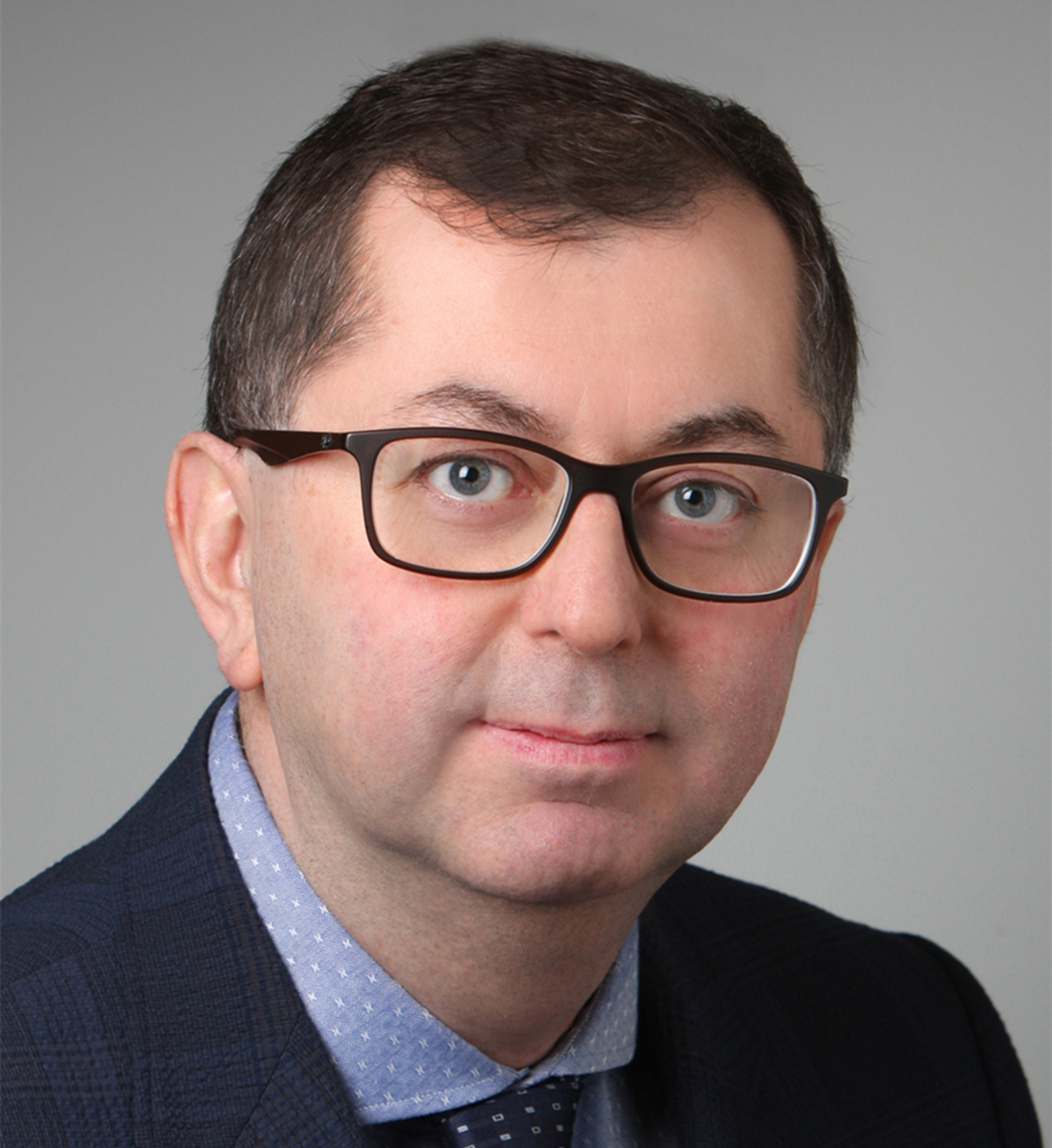}}]{Halim Yanikomeroglu}
\:(Fellow, IEEE) received his Ph.D. degree in Electrical and Computer Engineering from the University of Toronto (Canada) in 1998. \par
He is currently a Full Professor at Carleton University. He contributed to 4G/5G technologies and non-terrestrial networks. His industrial collaborations have resulted in 39 granted patents. He has supervised or hosted in his lab a total of 140 postgraduate researchers. He has co-authored IEEE papers with faculty members in 80+ universities in 25 countries. \par
Dr. Yanikomeroglu is also a Fellow of IEEE, Engineering Institute of Canada, and Canadian Academy of Engineering, and an IEEE Distinguished Speaker for ComSoc and VTS. He is currently chairing the WCNC Steering Committee, and he is a Member of PIMRC Steering Committee and ComSoc Emerging Technologies Committee. He served as the General Chair of two VTCs and TP Chair of three WCNCs. He chaired ComSoc’s Technical Committee on Personal Communications. He received several awards, including ComSoc Wireless Communications TC Recognition Award (2018), VTS Stuart Meyer Memorial Award (2020), and ComSoc Fred W. Ellersick Prize (2021).\par
\end{IEEEbiography}

\begin{IEEEbiography}[{\includegraphics[width=1in,height=1.25in,clip,keepaspectratio]{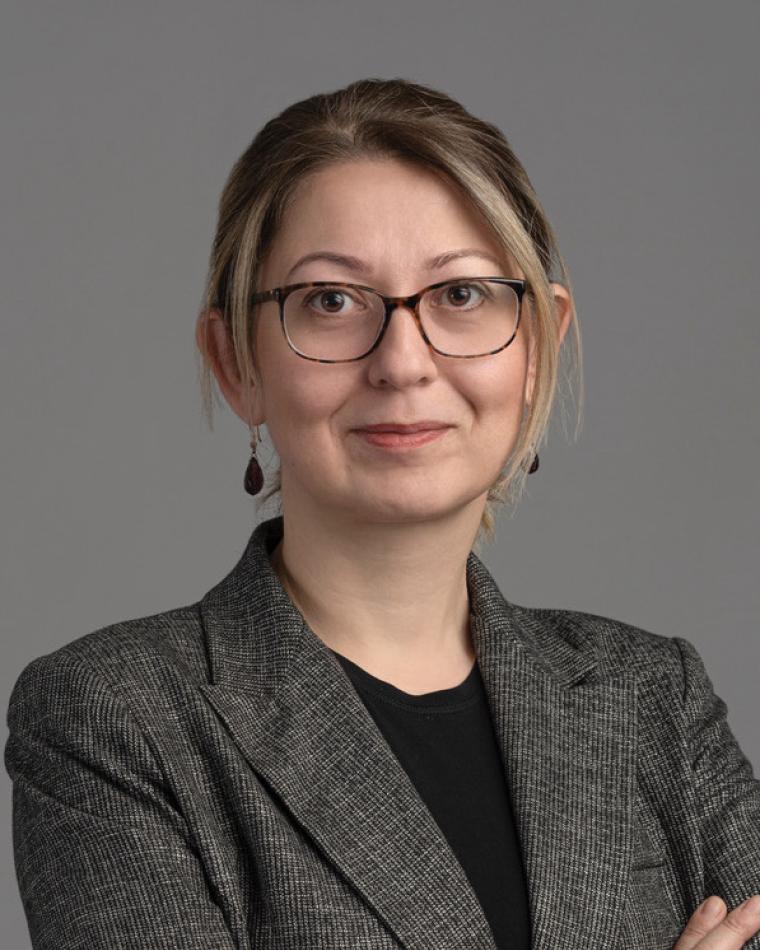}}]{Gunes Karabulut Kurt}
\:(Senior Member, IEEE) received the B.S. degree with high honors in electronics and electrical engineering from Bogazici University, Istanbul, Turkey, in 2000 and the M.A.Sc. and the Ph.D. degrees in electrical engineering from the University of Ottawa, ON, Canada, in 2002 and 2006, respectively. \par
From 2000 to 2005, she was a Research Assistant with the CASP Group, University of Ottawa. Between 2005 and 2006, she was with TenXc Wireless, Canada. From 2006 to 2008, Dr. Karabulut Kurt was with Edgewater Computer Systems Inc., Canada. From 2008 to 2010, she was with Turkcell Research and Development Applied Research and Technology, Istanbul. Between 2010 and 2021, she was with Istanbul Technical University. She is currently an Associate Professor of Electrical Engineering at Polytechnique Montréal, Montréal, QC, Canada. She is a Marie Curie Fellow and has received the Turkish Academy of Sciences Outstanding Young Scientist (TÜBA-GEBIP) Award in 2019. In addition, she is an adjunct research professor at Carleton University. \par
She is currently serving as an associate technical editor of the \textit{IEEE Communications Magazine}, an associate editor of \textit{IEEE Communication Letters}, an associate editor of \textit{IEEE Wireless Communications Letters}, and an area editor of \textit{IEEE Transactions on Machine Learning in Communications and Networking}. She is a member of the IEEE WCNC Steering Board. She is serving as the secretary of IEEE Satellite and Space Communications Technical Committee and also the chair of the IEEE special interest group entitled “Satellite Mega-constellations: Communications and Networking”. She is a Distinguished Lecturer of Vehicular Technology Society Class of 2022. \par
\end{IEEEbiography}

\begin{IEEEbiography}[{\includegraphics[width=1in,height=1.25in,clip,keepaspectratio]{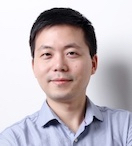}}]{Peng Hu}
\:(Senior Member, IEEE) received his Ph.D. degree in Electrical Engineering from Queen's University, Canada. \par
He is currently a Research Officer at the National Research Council Canada and an Adjunct Professor at the Cheriton School of Computer Science at the University of Waterloo. He has served as an Associate Editor of the IEEE Canadian Journal of Electrical and Computer Engineering, a voting member of the IEEE Sensors Standards committee, and on the organizing/technical committees of industry consortia and international conferences/workshops at IEEE ICC'23, IEEE PIMRC'17, IEEE AINA'15, etc. \par
His current research interests include satellite-terrestrial integrated networks, autonomous networking, and industrial Internet of Things systems. \par
\end{IEEEbiography}

\begin{IEEEbiography}[{\includegraphics[width=1in,height=1.25in,clip,keepaspectratio]{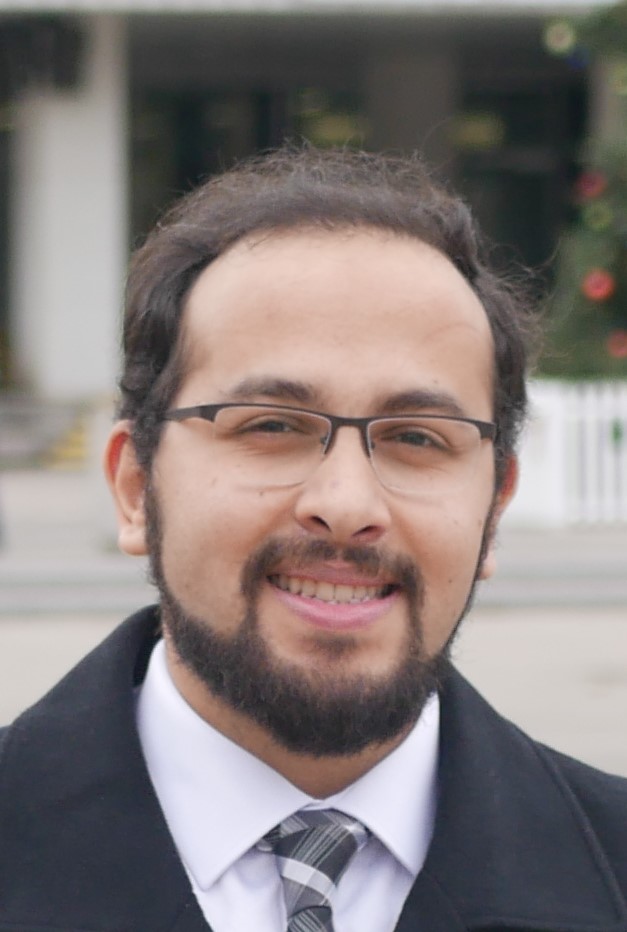}}]{Khaled Ahmed}
\:received his B.S. and MSc degree on wireless communications from Cairo university, Egypt, in 2010 and 2015. He did a PhD and a postdoc in McMaster university, Canada, in 2019 and 2021. His PhD degree was on optical communications while his postdoc was on machine learning for resource-limited devices. \par 
He is currently working as a Member Technical Staff II for MDA, Sainte-Anne-de Bellevue, QC, Canada. \par
His current research interests include RF and optical communications for terrestrial and satellite networks, optical beamforming networks, and applied machine learning and deep learning. \par
\end{IEEEbiography}

\begin{IEEEbiography}[{\includegraphics[width=1in,height=1.25in,clip,keepaspectratio]{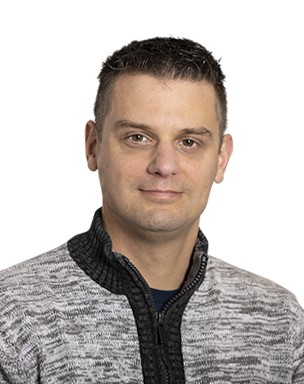}}]{Stéphane Martel}
\:received his Bachelor’s degree in Electrical Engineering from McGill University in 2001.\par 
He worked until 2021 in the broadcast television industry where he occupied multiple positions in software and management. He specializes in network communications and embedded software development. \par
He joined MDA Satellite systems in 2021 as R\&D product development manager where he contributes to the advancement of space communication technologies.\par
\end{IEEEbiography}
\end{document}